\documentclass{statsoc}

\usepackage{natbib}
\usepackage{cite}
\usepackage{ctable}
\usepackage{amssymb}

\renewcommand{\theenumi}{A\arabic{enumi}}

\newtheorem{thm}{Theorem}
\newtheorem{lma}{Lemma}

\title[Improved ranking and selection]{Making the cut: improved ranking and selection for large-scale inference }

\coaddress{Michael Newton; Department of Statistics, University of Wisconsin, 1300 University Ave, 
   Madison, WI 53706, USA.}

\author[Henderson {\it et al.}]{Nicholas C. Henderson}
\address{Department of Statistics, University of Wisconsin, Madison, USA.}
\email{nhenders@stat.wisc.edu}

\author[MAN]{Michael A. Newton}
\address{Departments of Statistics and of Biostatistics and Medical Informatics, 
 University of Wisconsin, Madison, USA.}

\email{newton@stat.wisc.edu}
\tr{1175.v7, UW Department of Statistics, June 22, 2015 }
\runauthor{Henderson and Newton}

\begin{document}

\begin{abstract}
Identifying leading measurement units from a large collection is a common
inference task in various domains of large-scale inference.  
Testing approaches, which measure evidence against a null hypothesis rather than
effect magnitude, tend to overpopulate lists of leading units with those
associated with low measurement error.  By contrast, local maximum likelihood (ML)
approaches tend to favor units with high measurement error.  Available
Bayesian and empirical Bayesian approaches rely on specialized
 loss functions 
that result in similar deficiencies.  
We describe and evaluate a generic empirical
Bayesian ranking procedure that populates the list of top units in a way
that maximizes the expected overlap 
between the true and reported top lists for all list sizes.
The procedure relates unit-specific posterior upper tail probabilities with their empirical
distribution to yield a ranking variable. 
 It discounts high-variance units less than popular non-ML methods
and thus  achieves improved operating characteristics in the models considered.
\end{abstract}
\maketitle

\section{Introduction}

In all sorts of applications,  data from a large number of  measurement or inference units
are processed in order to identify the most important units by some measure.  This is certainly true in 
statistical genomics, where units might be genes, gene sets,  or single-nucleotide 
polymorphisms (SNPs),
depending on the particular application, but it is also true more broadly. In agriculture
investigators rank animals or plants by their breeding value ({\em e.g.}, de los Campos {\em et al.} 2013); performance
evaluations in health and social sciences are common ({\em e.g.}, Paddock and Louis, 2011).   
Typically, units are associated
 with unobserved real-valued parameters, and the  importance of each unit is linked 
 to the value of its parameter.  
 A case we consider is a  genome wide association study examining 
risk factors for type 2 diabetes (T2D), 
 in which the inference unit is the SNP, and the parameter of interest is a log-odds ratio measuring
the effect on disease probability of SNP genotype (Morris {\em et al.}, 2012). 
A second case  involves gene-set enrichment
among human genes that have been determined via RNA interference (RNAi) experiments 
 to affect influenza-virus replication (Hao {\em et al.}, 2013).
Units here are sets of genes annotated to particular biological functions and parameters measure enrichment levels.
We develop two further examples to exercise the statistical issues: one from sports statistics (units are basketball players), and
one from gene expression analysis (units are genes).
Had their been no measurement error we would summarize each case by ranking units according to values of their parameters, focusing
on the top of this list for further study.  We consider here the inference task to perform such ranking and selection from data.
Where the emphasis of large-scale inference has been testing in 
relatively sparse settings
({\em e.g.}, Efron, 2010), the present work addresses the 
 inference task to rank order non-null parameters when the
 signal is  relatively non-sparse.

A  natural ranking is  obtained by separately 
 estimating unit-specific parameters, for instance  by maximum
likelihood applied locally to each unit.  Since sampling fluctuations 
 more easily put high-variance units 
into the tails,  units associated with relatively high standard error are 
over-represented among the top units by this MLE ranking.  Another commonly used procedure comes from
 large-scale hypothesis testing, where
 units are ranked by their p-value relative to a reference null hypothesis.
Units associated with relatively low standard error are 
over-represented among the top units by this ranking since both effect size and standard error affect testing power. 
 Standard error in the T2D case is affected by various factors including
 SNP allele frequency;  set size affects standard error in the RNAi case. 
When there is little variation among
 unit-specific standard errors, the different approaches give essentially the same assessment
of the
  most important units.  However in many cases there is substantial variation in these standard errors,
 and quite different rankings can emerge. 

For contemporary large-scale applications, the 
classical theory of ranking and selection leaves much to be desired. It addresses sampling
probabilities like, ``under such-and-such a configuration of parameters and for sufficient amounts of data per unit
the probability exceeds such-and-such that the true  top $j$ units are among the observed top $k$  units''
({\em e.g.}, Gibbons, Olkin, and Sobel, 1979).   While relevant to some tasks, these probabilities are difficult to work with 
and the resulting procedures are not often used in applied statistics.  Theory is available on the sampling 
characteristics of empirical rankings ({\em e.g.}, Xie {\em et al.}, 2009; Hall and Miller, 2010).
Arguably, the thrust of methodological development for ranking and selection involves
hierarchical modeling coupled with Bayes or empirical Bayes inference.  Seminal contributions by Berger and Deeley (1988) and Laird and Louis (1989)
helped to establish a framework that covers many contemporary applications and that has been elaborated in important ways ({\em e.g.}, Shen and Louis, 1998;
Gelman and Price 1999; Wright, Stern, and Cressie 2003; Lin {\em et al.} 2006; Brijs {\em et al.} 2007; Noma {\em et al.}, 2010).
 We further elaborate
 this framework in an effort to 
 provide a more effective generic method for large-scale inference, 
 especially when 
 large-parameter units are in focus, when there are a lot of units, and
  when there is substantial variation in unit-specific standard errors.

Sampling artifacts of MLE and p-value ranking procedures, noted above, are well documented, 
but other approaches are also deficient. The insightful analysis of Gelman and Price (1999)
illustrates the difficulties and confirms that the common practice to rank by posterior expected value
suffers from the same artifact as the p-value ranking, namely that units associated with small posterior
standard deviation are over-represented on lists of the top units.  We find similar behaviour with
the posterior expected rank method 
(Laird and Louis, 1989; Lin {\em et al.} 2006))
as well as available testing schemes.   
We introduce and investigate a procedure that aims to 
rank units in a way to maximize the expected overlap between the reported
 and the true top lists of units.
 While not eliminating the sampling artifacts, the new method reduces their effects compared
to other schemes. Our development starts in a special case wherein
ranking procedures are formulated in terms of certain threshold
functions (Section 2.1); using this formulation we characterize 
thresholds that maximize the expected overlap between the true and 
reported top lists (Section 2.2, 2.3), and we derive the associated
ranking variable in terms of local posterior tail probabilities.
The proposed {\em r-value} is generalized in Section 2.4 and investigated
in relation to other procedures in Section~3.  Computational issues are
reviewed in Section 4, sampling performance is investigated in Section~5, and
a short discussion follows.  Examples are used throughout for demonstration, and proofs are postponed until Section~7.
The proposed methodology and several data sets are deployed in 
the R package \verb+rvalues+, which is available through the Comprehensive
 R Archive Network (\verb+http://cran.r-project.org+).

\section{Threshold functions and ranking variables }
\subsection{Continuous model}
A variety of data structures are amenable to our proposed ranking/selection 
scheme, but the following
structure has guided its initial development.
  Measurement/inference units are indexed by $i=1,2, \ldots, n$; data on unit
$i$ include the real-valued 
measurement $X_i$ and information about its sampling variation.  We assume that
the sampling distribution of $X_i$ has a known form that is indexed by
 an unknown real-valued parameter of interest $\theta_i$ together with a second quantity affecting
variance.  In this section we
assume that $\sigma_i^2 = {\mbox {\rm var}}(X_i)$ is known for each unit. Basically, the inference task is
to report units having large values of $\theta_i$, while accounting for the fact that
variances $\sigma_i^2$ may fluctuate substantially among inference units. 
We adopt an empirical Bayes
perspective and treat $\{ (\theta_i, \sigma_i^2) \}$ as draws from a population of parameters, and we
are motivated by data-analysis considerations to suppose initially that $\theta_i$ and $\sigma_i^2$ are 
independent in this population, say with densities $f(\theta)$ and $g(\sigma^2)$.  
The independence assumption is helpful for understanding artifacts of various ranking methods, 
but it is not essential to the methodology.
The empirical Bayesian uses the full data set to estimate the {\em prior} distributions $f(\theta)$ 
and $g(\sigma^2)$. Initially we ignore the estimation
error at this level, and focus on ranking units within the estimated 
population, though we take up the issue in Section~5 via simulation and
 asymptotic analysis.

 Relative to a single unit $i$,  $X_i$ might be
the maximum likelihood estimator of $\theta_i$, and $\sigma_i$ that estimator's standard error.
The independence assumption may be reasonable if some care has been taken in this local analysis,
for example, by variance-stabilizing transformation.  
Typically, the variance $\sigma_i^2$ is estimated rather than known exactly; we study this
 and extensions to other data structures in Section~2.4. 
  We consider first a continuous model, involving prior 
distributions  and sampling distributions all having densities
with  respect to Lebesgue measure. 
 The canonical sampling model within this class has
$X_i | \theta_i, \sigma_i^2 \sim {\mbox {\rm Normal}}(\theta_i, \sigma_i^2 )$.  

We make some headway by associating each ranking/selection procedure with
a family $\mathcal{T}$ of threshold functions 
 $\mathcal{T} = \{ t_\alpha: \alpha \in (0,1) \}$. Each $t_\alpha$ is a function $t_\alpha( \sigma^2 )$
having the interpretation that unit $i$ is reported to be in the top $\alpha$ fraction of units
if and only if $X_i \geq t_\alpha( \sigma_i^2 )$. This interpretation is supported by the {\em size constraint}, namely,
that marginal to all parameters and data, 
\begin{eqnarray}
\label{eq:size}
P\left\{  X_i \geq t_\alpha(\sigma_i^2) \right\} = \alpha \quad {\mbox {\rm
		for all $\alpha \in (0,1)$ }}. 
\end{eqnarray}
Table~1 reports threshold functions associated with a variety of ranking methods
in the normal observation model, 
and under the extra condition that the  prior
$f(\theta)$ is Normal$(\mu, \tau^2)$.  The table encodes the special case $(\mu, \tau^2) = (0,1)$; 
the general thresholds are derived from this case by the transformation $\mu + \tau t_\alpha( \sigma^2/\tau^2 )$.
Note that each threshold function involves an $\alpha-$specific value $u_\alpha$ which guarantees the 
size constraint; these values are different for different ranking methods (rows of Table 1).
Figure~\ref{fig:T2D2} illustrates four of these families in the T2D case study. 
 Notionally, the linear ranking of units is obtained by sweeping through the family $\mathcal{T}$,
beginning with the smallest $\alpha$ at the top of the graph.   
  Clearly, distinct families of threshold functions can produce distinct rankings of the units, with  
the family's shape revealing how it
trades off observed signal $X_i$ with measurement variance $\sigma_i^2$ to prioritize the leading units.

Some comments on the threshold functions in Table~1 are warranted (see also
 the Supplementary Material document).  Under squared error loss, the Bayes estimate of the rank of
parameter $\theta_i$ among those in play is the conditional expected rank given the data (Laird and Louis, 1989; Lin {\em et al.} 2006).
This posterior expected rank (PER) is usually expressed as a sum, 
involving indicator comparisons between $\theta_i$ and the other parameters, and
it becomes $P(\theta_i \leq \theta | X_i, \sigma_i^2 )$ 
 when normalized and considered in the limit for increasing numbers of units (ranking from the top). 
Here $\theta$ is the independently drawn parameter of a generic additional unit, which emerges in the large-scale 
limit to replace the collection of all other $\theta_j$'s to which $\theta_i$ is compared. In the normal/normal model, 
ranking by posterior expected
rank is qualitatively similar to 
ranking by posterior mean PM = $E(\theta_i|X_i,\sigma_i^2)$; both favor small variance units.  
 Several hypothesis testing-based methods
are also shown in Table~1. 
 Testing against some benchmark null (rather than the no-effect null) has some benefits in practice
({\em e.g.}, McCarthy and Smyth, 2009).  As we emphasize large positive $\theta_i$, we report p-values (PV) 
associated with one-sided tests.  Finally, the BF entry aims to mimic the ranking (from the top) 
method associated with Bayes factors 
 for the test of $H_0: \theta_i  = 0$ versus $H_A: \theta_i \neq 0$ ({\em e.g.}, Kass and Raftery, 1995).
The mapping of a ranking method to a family of threshold functions is useful for comparative analyses, 
as we investigate next.

\subsection{Thresholds via direct optimization}

\begin{figure}
 \centering
 \makebox{\includegraphics{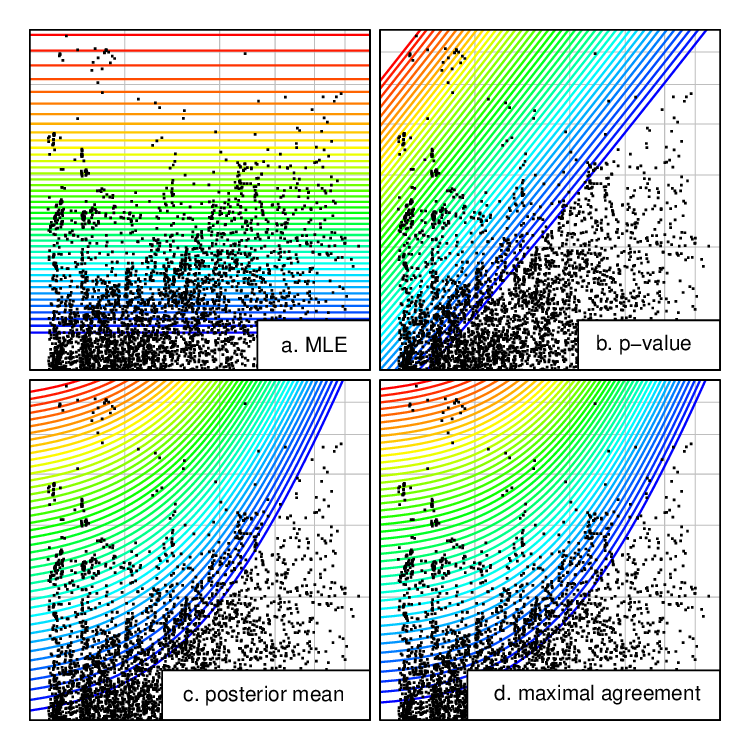}}
 \caption{Threshold functions, T2D example: Axes are common to all panels, with
  vertical the log odds ratio for association between SNPs (dots) and T2D, and with
 the horizontal the standard error estimates, with further details in Fig. S1, Supplementary Material:  
 Calculations use an inverse-gamma model for $\sigma^2$. Forty two 
 threshold functions are shown,
ranging in $\alpha$ values from a small positive value (red) just including the first 
 data point up to $\alpha= 0.10$ (blue). (Most SNPs are truncated by the plot;
  also, the grid is uniform on the scale of $\log_2 [ -\log_2 (\alpha) ]$.) 
   Units associated with a smaller $\alpha$ (i.e., more red) are ranked more highly
 by the given ranking method. Two units landing on the same curve would be ranked in the same position.
  \label{fig:T2D2} }
\end{figure}

 Table~1 and Figure~\ref{fig:T2D2} introduce a family
$\mathcal{T}^* = \{ t^*_\alpha \}$ that is optimal in the continuous model
in the sense that for all $\alpha \in (0,1)$:
\begin{eqnarray}
\label{eq:opt}
P \left\{  X_i \geq t^*_\alpha( \sigma_i^2 ) \, , \,  
 \theta_i \geq \theta_\alpha \right\} \geq
P\left\{  X_i \geq t_\alpha( \sigma_i^2 ) \, , \,  \theta_i 
 \geq \theta_\alpha \right\}
\end{eqnarray}
for any other family $\mathcal{T}= \{t_\alpha\}$ which also satisfies
the size constraint~(\ref{eq:size}).  Here $\theta_\alpha$ is the $\alpha$ upper
quantile of the prior; that is $P( \theta_i \geq \theta_\alpha ) = \alpha$.  In other words,
$\mathcal{T}^*$ maximizes {\em agreement}: the joint probability that unit $i$ is placed in the top $\alpha$ 
fraction  and its driving parameter $\theta_i$ is in the top $\alpha$ fraction of the
population, for all $\alpha$.  We emphasize that
the probabilities in~(\ref{eq:opt}) cover the joint distribution of  $X_i, \sigma_i^2, \theta_i$, 
which respects both the sampling distribution of data local to unit $i$ and the fluctuations of
unit-specific parameters.  
A calculus-of-variations argument provides
direct optimization of the joint probability in~(\ref{eq:opt}), subject 
to the size constraint, model regularity,  and 
smoothness of the threshold functions.
\begin{thm}
In the continuous model, a
 necessary condition for the function $t_\alpha^*$ to be optimal
as in~(\ref{eq:opt}), within the class of
continuously differentiable threshold functions, is that it satisfies:
\begin{eqnarray}
\label{eq:define}
P\left\{  \left. \theta_i \geq \theta_\alpha  \right| X_i = t^*_{\alpha}(\sigma^2), \sigma^2_i=\sigma^2 \right\}
 = c_\alpha \qquad {\mbox {\rm for all $\sigma^2$.}}
\end{eqnarray}
\end{thm}
Thus, all observations coincident with the graph of a given optimal threshold curve
have a common posterior probability $c_\alpha$ 
 that their unit-specific parameters
exceed the quantile $\theta_\alpha$  associated with that curve.
In the normal model for $X_i$ and the normal prior $f(\theta)$, 
the optimal threshold function (panel d, Figure~\ref{fig:T2D2}) is readily
extracted from~(\ref{eq:define}). Working on a standardized scale without loss of generality
 ($\mu=0$ and $\tau^2=1$), the local posterior for $\theta_i$
is normal with mean $X_i/(\sigma_i^2+1)$ and variance $\sigma_i^2/(\sigma_i^2 +1)$. Thus,
\begin{eqnarray}
\label{eq:tstar}
t^*_\alpha( \sigma^2 ) =  \theta_\alpha (\sigma^2 + 1 ) - u_\alpha \sqrt{ \sigma^2 (\sigma^2 + 1 ) },
\end{eqnarray}
where $\theta_\alpha = \Phi^{-1}(1-\alpha)$ and  $u_\alpha$  is determined by the size constraint~(\ref{eq:size}).
Indeed $u_\alpha$ is affected by the distribution $g(\sigma^2)$, since it is defined 
implicitly by: 
\begin{eqnarray}
\label{eq:ci}
1-\alpha = \int_0^\infty  \Phi\left\{  \theta_\alpha \sqrt{\sigma^2+1}  - u_\alpha \sigma  \right\} 
 \, g( \sigma^2 ) \, d\sigma^2
\end{eqnarray}
where $\Phi$ is the standard normal cumulative distribution function. 

Curiously, the optimal
thresholds {\em kick up} as $\sigma^2$ approaches zero.  
The resolution and range of Figure~\ref{fig:T2D2} do not reveal this phenomenon so clearly in the T2D example, but 
 it is apparent from~(\ref{eq:tstar}) that the derivative
of $t_\alpha^*$ with respect to $\sigma^2$ becomes increasingly negative as $\sigma^2$ approaches zero
 (when $u_\alpha > 0$).
 Neither the p-value thresholds nor those based upon posterior mean or 
posterior expected rank have this characteristic; indeed, by kicking up for small $\sigma^2$ the
maximal-agreement thresholds are less prone to the over-ranking of small-variance units.

\begin{figure}
 \centering
 \makebox{\includegraphics{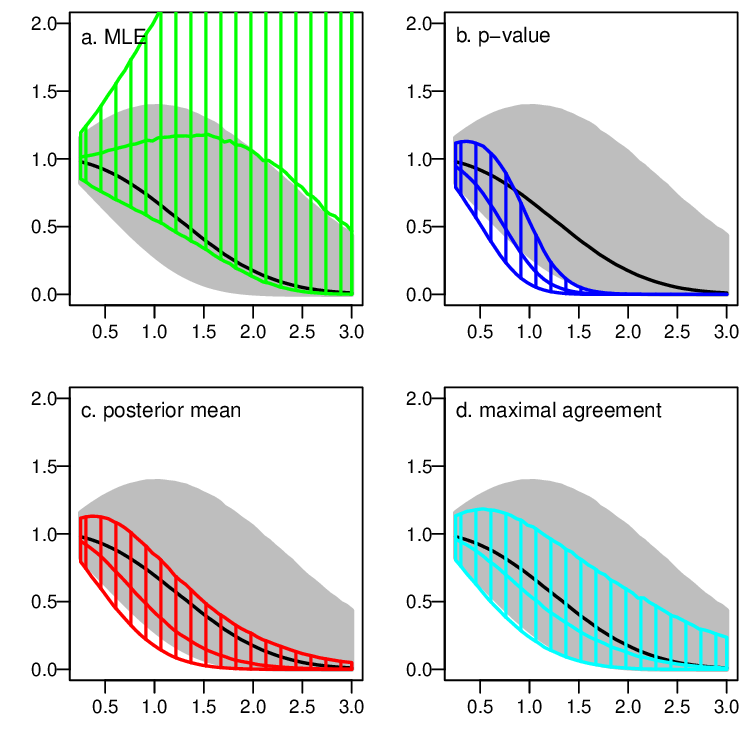}}
 \caption{ The conditional distribution (median, interquartile range) 
 of unit-specific variance $\sigma_i^2$ given selection of the unit
 in the top $\alpha=0.1$ fraction by various methods (colored bands) compared to 
 the marginal Gamma distribution (black, grey) for different amounts of variation in $\sigma_i^2$;
 $E(\sigma^2)=1$, coefficient of variation on the horizontal; 
based on simulation using $10^7$ units per case. \label{fig:vv} }
\end{figure}

Figure~\ref{fig:vv} illustrates 
 sampling properties of top-listed units obtained by various threshold schemes, including 
the optimal threshold~(\ref{eq:tstar}), using the normal/normal model, $\alpha = 0.1$, a sequence of
 Gamma distributions $g$ for the variance $\sigma_i^2$, and independence
between $\theta_i$ and $\sigma_i^2$. 
 The difference between different methods becomes more
pronounced as we increase variation in the distribution of the variances;  in this simulation all cases involve
 $E(\sigma_i^2) =1 $, but the shape parameters vary in order to increase the coefficient of variation.
The degree to which the conditional distribution of $\sigma_i^2$ given placement on the top $\alpha$ list
(colored bars) differs from the marginal distribution of $\sigma_i^2$ in the system (grey) measures 
the extent of sampling artifacts by that method.  
The example recapitulates sampling artifacts of the local MLE, 
the p-value, and the posterior mean. For example, the top lists by MLE are enriched for high-variance units.  The figure
also  shows that this artifact is substantially reduced when we select by~(\ref{eq:tstar}).

\subsection{Posterior tail probabilities and ranking variables}

Except in stylized models we cannot solve equation~(\ref{eq:define}) to identify optimal thresholds for ranking.
Insight into their structure comes by further
examining their relationship to local posterior tail probabilities:
$V_\alpha(X_i, \sigma_i^2) = P( \theta_i \geq \theta_\alpha | X_i, \sigma_i^2 )$.

\begin{thm}
Suppose that for $\alpha \in (0,1)$ there exists $\lambda_\alpha$ such that
\begin{eqnarray}
\label{eq:lambda}
P\left\{ V_\alpha( X_i, \sigma_i^2 ) \geq \lambda_\alpha \right\} = \alpha,
\end{eqnarray}
and furthermore that $V_\alpha(x,\sigma^2)$ is 
right-continuous and non-decreasing in $x$ for fixed $\alpha$ and 
$\sigma^2$. Then the family of thresholds,
$t^*_\alpha(\sigma^2) = \inf \{ x: V_\alpha(x, \sigma^2) \geq \lambda_\alpha \}$,
satisfies the size constraint~(\ref{eq:size})
 and is optimal in the sense of~(\ref{eq:opt}).
\end{thm}
The conditions above concern $V_\alpha$ and its distribution. They are satisfied in 
the normal/normal model; there, $V_\alpha(x,\sigma^2) = 1 - \Phi\left[ \sqrt{ \frac{\sigma^2 + 1}{\sigma^2}}
 \left( \theta_\alpha - \frac{x}{\sigma^2+1} \right) \right]$ and 
 $\lambda_\alpha = 1-\Phi(u_\alpha)$, for $u_\alpha$ is as in~(\ref{eq:ci}).
  The conditions are also satisfied in other instances of the  continuous model of Section~2.1, as well
as in other settings.  For example, if $\sigma_i^2$ is an estimated variance, then a Student~$t$ sampling 
model might replace the normal sampling model conditional on $\sigma_i^2$
and $\theta_i$.  See Supplementary Material for this and other examples.
Note that the optimal threshold $t_\alpha^*(\sigma^2)$ in Theorem~2 simplifies further
if $V_\alpha(x,\sigma^2)$ is continuous and strictly increasing in $x$ for each
$\alpha$ and $\sigma^2$.  Then $t_\alpha^*(\sigma^2) = V_\alpha^{-1}\left( \lambda_\alpha,
 \sigma^2\right)$, with the inverse referring to the first ({\em i.e.}, $x$) argument.

A family of threshold functions is a  device to think about 
converting observations into rankings ({\em i.e.}, by  
sweeping through the family).
Indeed, the index $\alpha$ associated with the threshold curve on which
data point $(X_i, \sigma_i^2)$ lands is a ranking variable;
its computation amounts to solving the inversion $X_i = t_\alpha( \sigma_i^2 )$
for $\alpha$. 
Exact inversion is possible as long as the threshold curves for different $\alpha$ values
do not cross: i.e., if there are no values $\alpha_1 < \alpha_2$, $\sigma^2$
 for which $t_{\alpha_1}(\sigma^2) = t_{\alpha_2}(\sigma^2)$.
  Approximate inversion is always possible via $\inf\{ \alpha: X_i \geq t_\alpha(\sigma^2) \}$.
\begin{thm}
Suppose that threshold functions $t_\alpha(\sigma^2)$ are differentiable
in $\alpha$ for each $\sigma^2$.
No functions in the family cross as long as 
 $\frac{\partial}{\partial \alpha} t_\alpha( \sigma^2 ) < 0 $
 for every $\alpha \in (0,1)$. 
Further, the optimal thresholds in the  normal/normal model do not cross.
\end{thm}
This confirms more generally what we see empirically for a few cases
in Figure~\ref{fig:T2D2} and Table~1: the  optimal thresholds
do not cross under the conditions of Theorem~3, and they conform to our intuition about
how ranking procedures might be constructed from threshold functions. 

We introduce a special ranking variable that inverts the optimal 
threshold.  For the $i$th unit, we define the {\em r-value}:
\begin{eqnarray}
\label{eq:rvalue}
r(X_i, \sigma_i^2 ) = \inf \left\{ \alpha:  V_\alpha( X_i, \sigma_i^2) \geq \lambda_\alpha 
		\right\}.
\end{eqnarray}
Essentially, unit $i$ is placed  by its r-value at position $\alpha$ (a relative rank, 
 measured from the top) if 
when ranking the units by $V_\alpha(X_i, \sigma_i^2)$, it also happens to land
at position $\alpha$.  Further, the top $\alpha$ fraction of units by
 r-value has higher overlap with the true top $\alpha$ fraction of units  than could be obtained by any other ranking procedure, in the sense 
 of~(\ref{eq:opt}).

It is worth recognizing that these findings go beyond what has been reported
about the use of the conditional tail probability
 $V_\alpha(X_i, \sigma_i^2)$ to rank units. Classical theory
on optimal selection establishes the role of this conditional tail probability
 in maximizing an exceedance probability within the selected sample ({\em e.g.}, Lehmann, 1986,
pages 117-118). Also, the conditional tail probability has been used
for ranking ({\em e.g.}, Normand {\em et al.}, 1997; Niemi, 2010), and is closely related to a Bayes optimal
ranking under a certain loss function (Lin {\em et al.}, 2006). A critical 
difference with the proposed ranking is in the role of the index $\alpha$. Conceptually,
we imagine ranking the units by $V_\alpha(X_i, \sigma_i^2)$ 
 separately for all possible indices $\alpha$ (not just a pre-specified one); 
then the r-value for unit $i$ is the smallest index $\alpha$ such that unit $i$
is placed in the top $\alpha$ fraction by that ranking. By aiming to
 maximize agreement at all list sizes, the proposed method does not require
 a pre-specified exceedance level to generate its  ranking.

\subsection{More generality}
The r-value construction makes sense in various elaborations of the 
model from Section 2.1.  We retain univariate parameters of interest $\{ \theta_i \}$ varying according to a distribution $F$, 
but we allow data $D_i$ on each unit to take more general forms than the $(X_i, \sigma_i^2)$ pair structure. 
We also retain the assumption of 
 mutual independence among units, though extensions 
could be developed in cases where posterior
computation is feasible. In seeking units with largest $\theta_i$, the critical quantity is the local exceedance probability,
$V_\alpha(D_i) = P\left( \left. \theta_i \geq \theta_\alpha  \right| D_i \right)$,
for $\alpha \in (0,1)$ and for upper quantiles $\theta_\alpha$ of the marginal distribution $F$: {\em i.e.},
$\theta_\alpha = F^{-1}(1-\alpha)$. Induced by the marginal distribution of $D_i$, the tail probability
$V_\alpha(D_i)$ has cumulative 
distribution function $H_\alpha(v)$, and from it we obtain the upper quantile:
$\lambda_\alpha = H_\alpha^{-1}( 1-\alpha )$.   Then by analogy to~(\ref{eq:rvalue}),  the r-value is 
 defined:
$r(D_i) = \inf \left\{ \alpha:  V_\alpha( D_i ) \geq \lambda_\alpha  \right\}$.

\begin{figure}
 \centering
 \makebox{\includegraphics{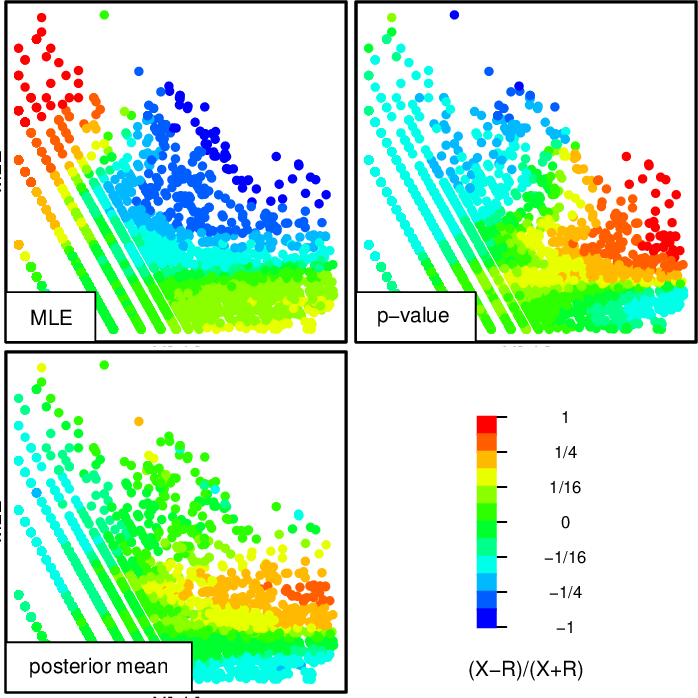}}
 \caption{Ranking via various methods compared to r-value ranking,
  RNAi example: Data and axes are common to all panels, with further details in Fig. S2, Supplementary Material.  Briefly,
 the horizontal axis is set size (log-scale) and the vertical axis is gene set enrichment.
   Each set (dot) is colored by $(X-R)/(X+R)$ where $X$ is the rank (from the top) of the set by the method being compared, and $R$ is
   the rank by r-value.  \label{fig:RNAi2}  }
\end{figure}

Figure~\ref{fig:RNAi2} compares r-value rankings with three other methods in the RNAi example.
Here,
$D_i=(m_i, y_i)$ holds binomial information (set size $m_i$ and number $y_i$ of genes in set $i$ that were identified by RNAi). 
The target parameters $\theta_i$ are treated as draws from a Beta$(a,b)$ distribution, 
with shape parameters estimated by marginal maximum likelihood, and 
the conditional tail probability $V_\alpha(D_i) $ becomes the 
 probability that a Beta$(\hat a+y_i, \hat b+m_i - y_i)$ variable exceeds $\theta_\alpha$.
R-value computation (see Section 4) requires the sampling distribution of these tail probabilities, which we approximated using the data
from all 5719 sets under study.  The methods compared in Figure~\ref{fig:RNAi2} agree to some extent on the ranking of the most 
interesting sets, but 
 systematic differences are apparent.  Ranking by $y_i/m_i$ over-ranks small sets; ranking by p-value over-ranks large sets; and ranking
 by posterior mean $(y_i+\hat a)/(m_i+\hat a+\hat b)$ also over-ranks large sets, though to a lesser degree, all compared to the r-value ranking.  

Sports enthusiasts routinely rank players. To explore r-value ranking in this context, 
  we deploy the same  Beta-Binomial model used in the RNAi example 
and use it to describe free-throw statistics of professional basketball players (e.g., Richey and Zorn, 2005). 
During the 2013-2014 regular season of the National Basketball Association (NBA), 461 players attempted at least one free throw
(ESPN, 2014). In total these players attempted 58,029 free throws and were successful 43,870 times, for a marginal
free-throw percentage of $75.6 \%$.  A basic problem in rating players by individual free-throw  percentage ${\rm FTP} = y_i/m_i$ is that the
numbers  $\{m_i\}$ of free-throw attempts vary substantially among players; in retaining all active players, those with highest $y_i/m_i$
are among those with smallest $m_i$. For instance, $13$ of the $461$ NBA players had perfect free-throw records in 2013-2014; they had
 a median number of $4$ attempts, compared to the league median of 82 attempts. Various threshold schemes have been adopted by rating agencies;
these restrict ranking to players reaching a minimum number of attempts or a minimum number of makes.  At ESPN, a {\em qualified} player this last
season needed $y_i \geq 125$.  Thresholding rules have a practical appeal but they can suppress athletic performances that otherwise are
 exceptional and worth reporting. For instance, Ray Allen's 105 makes in 116 attempts is exceptionally good by many standards (Table~\ref{tab:NBA}).
The context provided by the NBA example offers further insights.  For one thing, there is broad agreement between posterior mean (PM) ranking
 and r-value ranking, though where there is disagreement the PM favors players having more attempts $m_i$.
Related to this is the fact that though it discounts players with very small $m_i$, the r-value shrinks less than PM and is more in accordance
with the FTP ranking; for example, r-value ranks  the qualified players in Table~\ref{tab:NBA} the same as FTP, in contrast to PM.

As an empirical validation of the r-value ranking we applied it to mid-season NBA data (up to end of December, 2013),
 and then measured its performance conditional upon complete season data. Comparing Table~S2 (Supplementary Material) 
with Table~\ref{tab:NBA}, we
see some interesting features. For example, Brian Roberts, who finished the season with the highest FTP among qualified 
players, did not miss
in 2013; r-value placed him 2nd mid-season, even though he had only $m_i=18$ attempts, where PM ranked him 12th.  Investigating more fully,
we repeatedly simulated $\{ \theta_i \}$ vectors conditional upon end-of-season
 data, and averaged a similarity score: $\frac{1}{t} \sum_{i=1}^t 1[ {\rm rank}(\theta_i) \leq t ]\, 1[ \hat {\rm rank}_i  \leq t  ] $,
finding improvements over FTP and PM in assessing the best free-throw shooters (Figure~S3, Supplementary Material). 
 Here $\hat {\rm rank}_i$ is
 the player's estimated rank according to mid-season data and ${\rm rank}(\theta_i)$ is his unknown true rank.

R-values may be computed in all sorts of hierarchical modeling efforts, including semiparametric models and cases where Markov 
 chain Monte Carlo (MCMC) is
used to approximate the marginal posterior distribution of each $\theta_i$ given available data.  Figure~S9 (Supplementary Material) 
 compares the r-value ranking
with other rankings in an example from gene-expression analysis, where evidence suggested that the expression 
of  a large fraction of the human genome was associated with the status of a certain viral infection (Pyeon, {\em et al.}, 2007).
A multi-level model involving both null and non-null genes as well as $t-$distributed non-null effects $\theta_i$ exhibited good fit to the data, 
but did not admit a closed form for $V_\alpha(D_i)$. 
 R-values, computed using MCMC output, again reveal systematic ranking differences from other approaches. 

Multi-level models drive statistical inference and software in a variety of genomic domains: for example,  \verb+limma+ (Smyth, 2004),
\verb+EBarrays+ (Kendziorski {\em et al.} 2003), \verb+EBSeq+ (Leng {\em et al.} 2013), among others. Since these models 
happen to specify distributional forms for parameters of interest, the associated code could be augmented to compute
posterior tail probabilities $V_\alpha(D_i)$ and thus r-values for ranking. 
 The \verb+limma+ system utilizes a conjugate normal, inverse-gamma model,
and so $V_\alpha(D_i)$ involves the tail probability of a non-central $t$ distribution.  The \verb+EBSeq+ system entails a conjugate
beta, negative-binomial model, and so $V_\alpha(D_i)$ for differential expression involves tail probabilities in a certain ratio
distribution (Coelho and Mexia, 2007). One expects the benefits of r-value computation to show especially in cases involving many non-null units 
and relatively high variation among units in their variance parameters ({\em e.g.}, sequence read depth).
 The data structure envisioned for r-value computation involves many
exchangeable units, with real-valued parameters driving the conditional distribution of data on each unit.  
 Other structures, such as from large-scale
 regression, may be amenable to the proposed ranking method
  if marginal posterior distributions for each regression coefficient could be derived.

\section{Connections}
\subsection{Connection to Bayes rule}

The proposed r-values are not Bayes rules in the usual sense, however
there is a connection to Bayesian inference if one allows both a continuum of loss functions 
and a distributional constraint on the reported unit-specific (relative) ranks.
 To see this connection, we introduce a collection of loss functions
\begin{eqnarray*}
L_\alpha(a, \theta_i) = 1 - 1\left( a \leq \alpha, \theta_i \geq \theta_\alpha
	\right)
\end{eqnarray*}
where action $a$ is a relative rank
 value in $(0,1)$, $\alpha \in (0,1)$
indexes the collection, and again $\theta_\alpha= F^{-1}(1-\alpha)$ is
a quantile in the population of interest.  
 Specifically, no $\alpha-$loss occurs if 
the inferred relative rank $a$ and the actual relative rank
 $1-F(\theta_i)$ both are less than $\alpha$.
The marginal (pre-posterior) Bayes risk of rule 
 $\delta (D_i)$ is
\begin{eqnarray}
\label{eq;risk}
{\mbox {\rm risk}}_{\alpha} = 1 - P\left\{ \delta (D_i) \leq \alpha\, , 
 \theta \geq \theta_\alpha  \right\},
\end{eqnarray}
which is one minus the agreement~(\ref{eq:opt}).  In the absence of
other considerations, the Bayes rule for loss $L_\alpha$ degenerates to
$\delta(D_i) = 0$. Degeneration is avoided if we
enforce on the reported rank the additional structure that it share
with the true relative rank $1-F(\theta_i)$  the property of 
 being  uniformly distributed over the population of units.  
Such a constrained Bayes rule then  minimizes  the modified objective function:
${\mbox {\rm risk}}_\alpha \, +  \, \gamma_\alpha  
 P\left\{  \delta( D_i) \leq \alpha \right\}$,
where $\gamma_\alpha$ is chosen to enforce the (marginal) size constraint
 $P\left\{  \delta( D_i ) \leq \alpha \right\} = \alpha$.

The constrained Bayes rule
is computed conditionally, per observed $D_i$,
  by minimizing the constraint-modified posterior expected loss (PEL)
\begin{eqnarray}
\label{eq:pel}
{\mbox {\rm PEL}}_\alpha &= &
 1 - P\left\{ \left. \delta(D_i) \leq \alpha\, , \, 
		\theta_i \geq \theta_\alpha \right| 
	D_i \right\}
 	+ \gamma_\alpha 1\left\{ \delta(D_i)
			\leq \alpha \right\} \\ \nonumber
 &=&  \left\{ \begin{array}{ll}
1- V_\alpha( D_i ) + \gamma_\alpha &
 	{\mbox {\rm if $\delta (D_i ) \leq \alpha $}}\\
	1 & {\mbox {\rm if $\delta(D_i ) > \alpha$ }} \\
	\end{array}
	\right.
\end{eqnarray}
where $V_\alpha(D_i)$ is the upper posterior probability
$ P( \left. \theta_i \geq \theta_\alpha \right| D_i ) $ 
 appearing in
Section 2.

Curiously, a rule minimizing ${\rm PEL}_{ \alpha}$ is not uniquely determined at a single $\alpha$,
since minimization in~(\ref{eq:pel}) requires only that
\begin{equation}
\delta(D_{i}) \leq \alpha \qquad \Longleftrightarrow \qquad  V_{\alpha}( D_{i} ) \geq \gamma_{\alpha}.
\label{eq:equiv}
\end{equation}
However, taking all losses together does fix a procedure. 
To see this, let $g(\alpha|D_{i}) = V_{\alpha}(D_{i}) - \gamma_{\alpha}$, and further
assume $g$ is continuous in $\alpha$. If $g(\alpha|D_{i})$ has only one root in $(0,1)$, 
then the procedure 
$\delta^{*}(D_{i}) = \inf\{ \alpha: V_{\alpha}(D_{i}) \geq \gamma_{\alpha} \}$
is a Bayes rule for any choice of $L_{\alpha}$, even though $\delta^{*}$ does not depend on any specific
choice of $\alpha$. This is because $\delta^{*}(D_{i}) \leq \alpha$ for all $\alpha$ such that $g(\alpha|D_{i}) \geq 0$,
and $\delta^{*}(D_{i}) > \alpha$ for all $\alpha$ such that $g(\alpha|D_{i}) < 0$.
If $g(\alpha|D_{i})$ does contain multiple roots (at least over a range of $D_{i}$ that has positive
probability), there will not be a procedure (i.e., a procedure which doesn't
depend on $\alpha$) which is a Bayes rule for any choice of $L_{\alpha}$. This is because
it will not be possible to construct a rule $\delta$ that satisfies (\ref{eq:equiv}) for
all values of $\alpha \in (0,1)$.
 The thresholds $\gamma_\alpha$ in~(\ref{eq:equiv})
are determined by the uniformity constraint, and we have
$\gamma_\alpha = H_\alpha^{-1}(1-\alpha)$, where $H_\alpha$ is
the marginal distribution of $V_\alpha(D_i)$, counting
all sources of variation, and so $\gamma_\alpha=\lambda_\alpha$ from the previous section. 
 In other words, the procedure obtained by 
this constrained, multi-loss Bayes calculation is equivalent to the r-value
introduced in Section~2.

Among the more popular loss-based ranking procedures is one via posterior expected rank (PER) (e.g., Laird and Louis, 1989;
 Noma {\em et al.} 2010).
Unit $i$'s value 
 becomes ${\rm PER}_i = P(\theta_i \leq \theta | D_i)$ after normalizing by the number of units and taking the large-scale
 limit. We find in numerical experiments that PER ranking
is relatively close to the ranking by posterior mean (PM), and in these experiments we use: 
 ${\mbox {\rm PER}}_i = 1-\int_0^1 V_{\alpha} (D_i) \, d\alpha $,
which can be established readily using a transformation of variables argument.

\subsection{Beyond p's and q's}

In testing a single hypothesis $H_0$, the sample space may be structured as a nested sequence
of subsets, $\left\{ \Gamma_\alpha : \alpha \in (0,1) \right\}$, 
say, such that rejection of a size $\alpha$ test is equivalent
to data $D$ landing in set ({\em i.e.}, rejection region) $\Gamma_\alpha$.  Then, the p-value of the test is 
$ p(D) = \inf\{ \alpha: D \in \Gamma_\alpha \}$.
Storey (2003) extended this idea to multiple testing and the positive false discovery rate with the
introduction of the q-value. Specifically, with another nested sequence $\{ \tilde \Gamma_\alpha: \alpha \in (0,1) \}$
indexed such that
$ P(H_0 | D \in \tilde \Gamma_\alpha ) = \alpha$,
the q-value is $q(D) = \inf\{ \alpha: D \in \tilde \Gamma_\alpha \}$.
Where p-values refer to the distribution of $D$ on $H_0$, and q-values the conditional probability of $H_0$ given sample information,
the proposed r-values refer to marginal probability over both unit-specific data and unit-specific parameters.  
 The size constraint~(\ref{eq:size}) corresponds to another sequence
of subsets, $\{ \check \Gamma_\alpha \}$, say, for which the marginal constraint holds: $P( D \in \check \Gamma_\alpha) = \alpha$.
Analogously, the r-value is $r(D) = \inf\{ \alpha: D \in \check \Gamma_\alpha \}$.  In principle an r-value could be defined for
any indexed ranking method, though we have reserved the definition for that method which maximizes 
 agreement~(\ref{eq:opt}).  Other connections to hypothesis testing are discussed in Supplementary Material.

\section{Computation}
 In Section~2.2 we focused on the model involving normality for both the measurement 
$X_i$ and the 
 latent parameter $\theta_i$.  The r-value is obtained by inverting~(\ref{eq:tstar}) to solve for $r$:
\begin{eqnarray}
\label{eq:norm}
 X_i = (\sigma_i^2+1)  \Phi^{-1}(1-r)  - u_r \sqrt{ \sigma_i^2 (\sigma_i^2 + 1 ) }
\end{eqnarray}
where $u_r$, defined through the size-constraint~(\ref{eq:ci}), is readily computed numerically.

\begin{figure}
 \centering
 \makebox{\includegraphics[scale=.85]{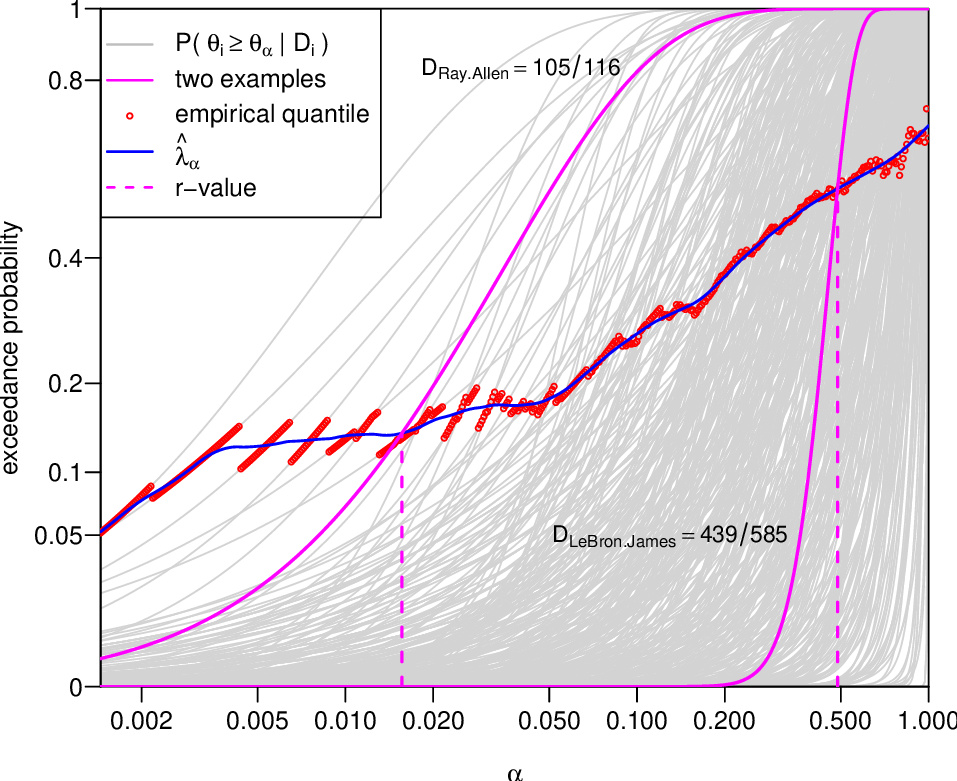}}
 \caption{Computational details, NBA example: Grey lines show, for each of
  461 NBA players who attempted at least one free throw in the entire 2013-2014 regular season,
  the tail probability function $V_\alpha(D_i) = P(\theta_i \geq \theta_\alpha| D_i )$; two are highlighted in magenta.
 Recall $\theta_\alpha$ is such that $P( \theta_i \geq \theta_\alpha ) = \alpha$; in this case a conjugate, Beta$(a,b)$ model was fit to obtain
 these marginal quantiles ($\hat a = 15.12, \hat b = 5.38$).  At each value of $\alpha[j]$ on a grid, the
  the empirical distribution of $\{ V_{\alpha[j]}(D_i) \}$ was computed and reduced to a quantile such that the empirical frequency
 exceeding the quantile is $\alpha[j]$ (red dots).
  We smoothed these to obtain the quantile function
  $\hat \lambda_\alpha$ (blue curve). Two r-values are shown
 (dashed lines, at r-values $0.016$ and $0.488$),
 obtained by solving in $\alpha$
 equality of the unit-specific $V_\alpha(D_i)$ and the system-wide $\hat \lambda_\alpha$.
 Scaling by log (horizontal) and square root (vertical)
 was done to aid visualization. \label{fig:NBA1} }
\end{figure}

Alternatively, a 
 generic approach to computing r-values starts with a finite grid $\{ \alpha_j \}$ in $(0,1)$, 
at which we compute the posterior tail probabilities 
$v_{i,j} = V_{\alpha_j}(D_i) $ for
all units $i$ (or approximations, {\em e.g.} by MCMC). The grid need not be uniform;
we enrich coverage near $0$ in our implementation. 
The $j$th column of the matrix $\{ v_{i,j} \}$ holds a sample from the marginal distribution
for which $\lambda_{\alpha_j}$ is the $1-\alpha_j$ quantile. Marching through $j$ allows us
to assemble a discrete (in $\alpha$), empirical (over units) quantile function, which we convert 
to a function $\hat \lambda_\alpha$ first by possibly smoothing to mitigate sampling effects and then 
by interpolating to $\alpha$ values beyond the initial grid.  Then for each unit $i$ we
solve $V_\alpha( D_i) = \hat \lambda_{\alpha}$ numerically in $\alpha$ to obtain
that unit's r-value.  Figure~\ref{fig:NBA1} illustrates the computation for two units in the NBA example.
Pseudo-code for the algorithm and elements of the \verb+R+ package implementation are given 
in the Supplementary Material document. 

\section{Sampling performance}

\begin{figure}
 \centering
 \makebox{\includegraphics[scale=.9]{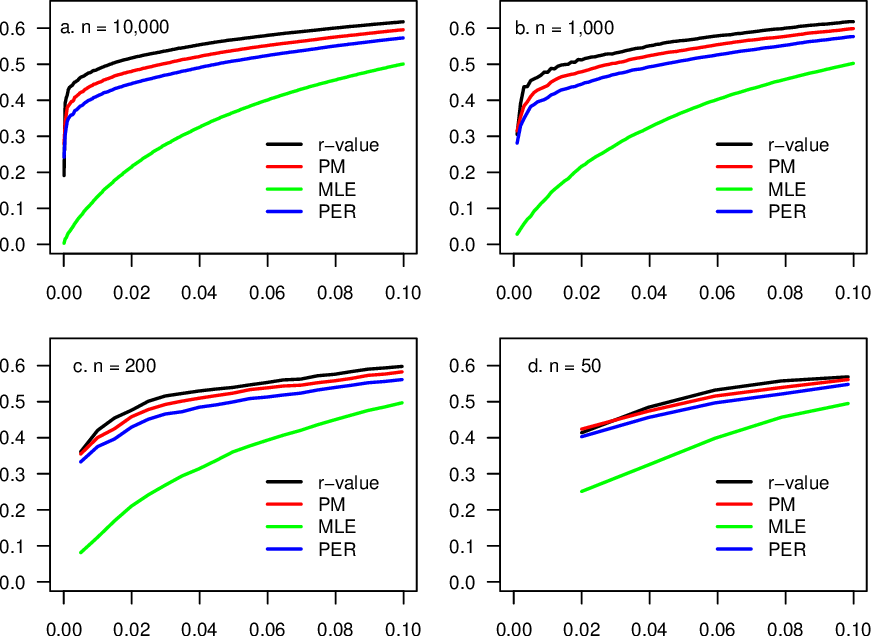}}
 \caption{ Finite-sample performance of r-value, posterior mean (PM),
 posterior expected rank (PER), and maximum likelihood estimate (MLE)
  in the normal/normal model.
 The simulation-based 
 agreement compares the true top-$\alpha$ list with the
 estimated top-$\alpha$   list for various methods and for
 $1/n \leq \alpha \leq 0.1$ (common horizontal axis), when
 the marginal distribution of $\theta_i$ and the quantile $\lambda_\alpha$ 
 are both estimated from available data (no smoothing).  The common
 vertical axis is agreement$/\alpha$; 
 $\sigma_i^2 \, \sim \, 
 {\mbox {\rm Gamma}}(1/2,1/2)$, and results from
  1000 simulated data sets were averaged for
  each panel.
 \label{fig:sim1}  }
\end{figure}

The r-value is defined 
 using the joint distribution of data $D_i$ and the target parameter $\theta_i$, 
 but it is computed empirically from an estimate of that joint distribution.  
 Accurate distributional estimation may 
 be possible from large-scale data sets, but
 it is nonetheless useful to investigate  how the optimality guaranteed by 
 Theorems~1 and~2 deteriorates in finite-sample situations.
  Simulations of the normal/normal model
  show that computed r-values 
  retain their performance benefits compared to other ranking procedures, 
 and thus some uncertainty in the quantile function $\lambda_\alpha$ or
 in the distribution of $\theta_i$ does not clearly disable the procedure.   For example,
 Figure~\ref{fig:sim1}
 shows simulation-based estimates of agreement,
  $P\left[  \hat r_n(D_i) \leq \alpha, \, 
                       \theta_i \geq \theta_\alpha \right] $,
 for both the computed r-values $\{ \hat r_n(D_i) \}$ and for other ranking
 methods.  We adapt the notation to  
 include the sample size $n$ and the {\em hat} mark
 in order to emphasize that the computed 
 r-values involve estimation of the marginal distribution function $F$ of $\theta_i$ and
 the quantile function $\lambda_\alpha$.
  R-value performance is not adversely affected
 by low sample sizes in this case.  Other simulations demonstrate that this
 superiority is not sensitive to the distribution of variances or
 to the extent of smoothing used to compute quantiles (see
 Supplementary Material, Figures S4 and S5). 

 A more general consistency property holds for models
sufficiently regular that the following four
conditions, A1-A4, are satisfied.
 \begin{enumerate}
 \item Triples $(\theta_i, X_i, \sigma_i^2)$,
 for $i=1,2,\ldots,n$, 
 are independent and identically distributed from a joint distribution
 for which: $\theta_i$ and $\sigma_i^2$  are independent and have positive
 densities $f$ and $g$ 
 with respect to Lebesgue measure on $\mathbb R$ and $\mathbb R^+$,
 respectively.  
 \item From data $\left\{ D_i=(X_i, \sigma_i^2): \, i=1,2, \ldots, n
     \right\}$, we have an estimator $\hat F_n$ of 
 	$F$, where $F(\theta)=\int_{-\infty}^\theta f(t) \, dt$, that
 is invariant under permutations of the observations.  The sequence of distributions
 converges weakly, $\hat F_n \Rightarrow F$,
 almost surely as $n \longrightarrow \infty$.  
 \end{enumerate}
The estimator $\hat F_n$ could be parametric or nonparametric ({\em c.f.}
 Lindsay, 1995).
For each $\alpha$, the marginal quantile $\theta_\alpha = F^{-1}(1-\alpha)$
 is estimated by 
 $\hat \theta_{\alpha ,n } = \hat F_n^{-1}(1-\alpha)$, and the 
 posterior tail probability, $V_\alpha(x,\sigma^2)$,
given a potential data point $(x,\sigma^2)$, is estimated by: 
\begin{eqnarray}
\label{eq:vhat}
\hat V_{\alpha, n}(x,\sigma^2) = 
  \int_{\hat \theta_{\alpha,n}}^\infty p(x|\theta,\sigma^2) \, d\hat F_n(\theta)   \Bigg/  \int_{-\infty}^\infty p(x|\theta,\sigma^2) \, d\hat F_n(\theta).
\end{eqnarray}
Here $p(x|\theta,\sigma^2)$ is the local sampling density, which we consider to 
have a known form. 
\begin{enumerate}
\setcounter{enumi}{2}
\item The local sampling density satisfies:
  \begin{enumerate} 
  \item  $p(x|\theta,\sigma^2)$ is continuous in $(x, \theta, \sigma^2)$,
  \item there is a continuous
 function $ K(\sigma^2) $ such that $0 < p(x|\theta,\sigma^2) \leq K(\sigma^2) $ for all arguments.
  \item for any $x_1 > x_0$ and $\sigma^2>0$, 
  $p(x_1|\theta,\sigma^2)/p(x_0|\theta,\sigma^2)$ is increasing in $\theta$.
   \end{enumerate}
\end{enumerate}
Let $H_\alpha(v) = P\left[ V_\alpha(X_i, \sigma_i^2) \leq v \right]$, 
  $\lambda_\alpha = H_\alpha^{-1}(1-\alpha)$, and 
  $t^*_{\alpha}(\sigma^2) = \inf\{ x: V_{\alpha}(x, \sigma^2) \geq \lambda_{\alpha} \}$.
\begin{enumerate}
\setcounter{enumi}{3}
\item There are no values of $\sigma^2$ and $\alpha_1 \neq \alpha_2$ such that $t^*_{\alpha_1}(\sigma^2) = t^*_{\alpha_2}(\sigma^2)$. 
\end{enumerate}
 The normal/normal model satisfies A1 by design, A3 by inspection, and A4 by Theorem~3, and will satisfy A2 for typical
parametric or nonparametric estimates of $F$.  Indeed A3 is readily verified in many settings, but A4 is more difficult because it involves the
marginal distribution of local posterior probabilities, which is often analytically intractable.
We have confirmed A1-A4 in a Gamma/Inverse-Gamma model (see Supplementary Material).

 The ideal r-value  $r(D_i) = \inf\{ \alpha \in (0,1): 
 V_\alpha(D_i) \geq \lambda_\alpha \}$ is not computable when the underlying distributions are unknown,
 though model regularity assures that  $r(D_i)$ is the unique root (in $\alpha$) of
 the equation $V_\alpha(D_i) = \lambda_\alpha$.
Approximating $H_\alpha(v)$ we
have the empirical distribution function,
$\hat H_{\alpha,n} (v) = \frac{1}{n} \sum_{i=1}^n 1\left[    \hat V_{\alpha, n}(X_i,\sigma_i^2) \leq v \right]$,
and the unsmoothed quantile $\hat \lambda_{\alpha,n} = \hat H_{\alpha,n}^{-1}(1-\alpha) = \inf\{ v: \hat H_{\alpha,n}(v) \geq 1-\alpha \}$.
 A natural estimate of $r(D_i)$ 
 is $\hat r_n(D_i) = \inf\{ \alpha \in (0,1): \hat V_{\alpha,n}(D_i) \geq \hat \lambda_{\alpha,n} \}$.
 To analyze estimation error,  it is helpful to define the related quantity
 $r^\delta(D_i) = \min\left[ \inf\{ \alpha \in [\delta,1]: V_\alpha(D_i) \geq \lambda_\alpha \}, 1- \delta \right]$
 for  $\delta \in (0,1/2)$, and the sample version,
 $\hat r^\delta_n(D_i) = \min\left[ \inf\{ \alpha \in [\delta,1]: \hat V_{\alpha,n} (D_i) \geq \hat \lambda_{\alpha,n} \}, 1- \delta \right]$.
It happens that $r^\delta(D_i) = r(D_i)$ when both reside in $[\delta, 1-\delta]$; we think of $\delta$ as an
arbitrarily small value that ameliorates boundary effects in the estimated quantile function $\hat H^{-1}_{\alpha,n}$.
\begin{thm}
If the model satisfies A1-A4 
 and $n \longrightarrow \infty$, then  for $\delta \in (0,1/2)$,  and all $\alpha \in [\delta, 1-\delta]$,
$ \frac{1}{n} \sum_{i=1}^n 1\left[  \hat r_n^\delta(D_i) \leq \alpha \right]
  \longrightarrow_P \alpha.$
Furthermore,
\begin{eqnarray}
\label{eq:converge2}
 \frac{1}{n} \sum_{i=1}^n 1\left[  \hat r_n(D_i) \leq \alpha , 
          \theta_i \geq \theta_\alpha \right]
  \, \geq \,  P\left[ r(D_i) \leq \alpha , \theta_i \geq \theta_\alpha 
 \right] + o_P(1).
\end{eqnarray}
The quantity $P\left[ r(D_i) \leq \alpha , \theta_i \geq \theta_\alpha \right]$  is the optimal agreement, as in Theorem~2.
\end{thm}
Essentially, computed r-values are uniformly distributed and achieve the maximal agreement in large samples as long as the generative distributions are sufficiently regular and consistently estimated.

\begin{figure}
 \centering
 \makebox{\includegraphics[scale=.9]{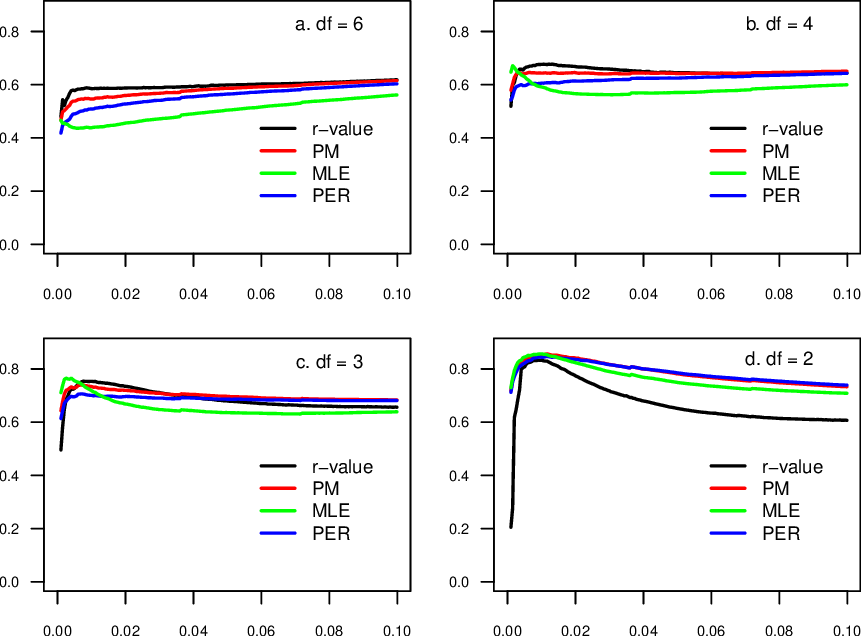}}
 \caption{ Effects on agreement of  model misspecification:
 R-value performance deteriorates when the true distribution of effects
 $\theta_i$ is much heavier tailed (Student's $t$ on $df$ degrees of freedom) 
 than is used to construct the r-value (normal). 
 The case shown involves $n=2000$. Axes and labels are as in Fig.~9.
 \label{fig:misspecify}  }
\end{figure}

Model uncertainty can have a bigger effect than system-parameter uncertainty  on the r-value performance.  Figure~\ref{fig:misspecify} shows 
some reduced
 performance of r-value in case $F$ is misspecified as 
 normal when it is fact heavier tailed.  Other misspecifications may 
have less effect, such as when the true $F$ is a finite mixture
 of normals, or
 when there are un-modeled dependencies between $\theta_i$ and $\sigma_i^2$.
 Examples are provided in the Supplementary Material, Figures S6-S8.
Without pursuing a comparative analysis, 
 we note finally that an alternative estimator of 
 $\lambda_\alpha = H^{-1}_\alpha (1-\alpha)$ may be obtained by 
 working out, perhaps via simulation, the induced distribution of $V_\alpha(D^*)$
 for bootstrap data $D^*$ drawn from the fitted model.

\section{Discussion}

For examples touched upon here as well as for many others within the domain of large-scale inference,
a basic statistical problem is to rank units and select the top ones by some measure.  Precisely how
the output of such inference is to be used depends very much on the context; admittedly we have not
focused on these operational issues.  For example, the output might trigger follow-up experiments in a genomic study (e.g., Pyeon {\em et al.} 2007),
it might affect resource allocation in some performance evaluation (e.g., Paddock and Louis, 2011); or it
might spark a debate about who really is the best free-throw shooter.  Our emphasis on a statistical framework for
large-scale ranking and selection responds to evident weaknesses of available methodologies and the potential
utility of the proposed r-value scheme, especially when there is great variation in the amount of information per unit.
Also, where an emphasis of large-scale inference has been on testing and sparsity assumptions, the r-value computation addresses
a practical problem to organize large numbers of non-null units.

By casting the problem via empirical Bayes, we express agreement
 between  true and reported top lists as a certain joint probability that is subject to explicit optimization, taking advantage of
an equivalence between ranking and threshold functions (Section 2).
Roughly speaking, an r-value is a Bayes rule for the binary loss which indicates failure to correctly place the unit
 into the top $\alpha$-fraction of units,
though to formalize this one requires multiple loss functions and a distributional constraint  (Section 3.1).
In spite of this connection
to Bayesian inference, the r-value method seems not to have been previously identified by that reasoning.
Theoretical support for the method has been developed here for a 
 measurement model (Sections 2.1-2.3).
Establishing that r-values maximize agreement in the more general cases
considered in Section 2.4 remains to be investigated. 
 Where the analysis in Section~2 treats the joint distribution of data and unit-level 
 parameters
 as known, this model must be
  estimated from system-wide data in each application. 
 We report sufficient conditions 
for  first-order asymptotic correctness (Theorem 4). 
 Within-model  simulations show good r-value performance under a range of conditions (Section 5).  Performance deteriorates
 when the model is misspecified, and we recommend that standard model diagnostics accompany the r-value computation. 
 Further investigation is warranted for nonparametric/semiparametric models, as the
 basic r-value statistic does not require a parametric formulation.

\section{Proofs}

\subsection{Theorem 1}  
In this section we assume that all distributions have continuous densities on their support.
From the calculus of variations ({\em e.g.}, Jost and Li-Jost, 1998, chapter 1),
for a continuously differentiable threshold function $t^*=t^*_\alpha( \sigma^2 )$ to maximize
agreement~(\ref{eq:opt}) subject to the size constraint~(\ref{eq:size}), it must be a critical
point of the objective function:
$I(t) = \int_0^\infty F(t, \sigma^2) \, g(\sigma^2)  \, d\sigma^2$,
where 
\begin{eqnarray*}
F(t, \sigma^2) = P\left\{ \left. X \geq t_\alpha(\sigma^2) , \theta \geq \theta_\alpha \right| 
 \sigma^2 \right\}
 + \lambda P\left\{ \left. X \geq t_\alpha(\sigma^2) \right| \sigma^2 \right\},
\end{eqnarray*}
and where $\lambda$ is a Lagrange multiplier.
Here and to follow we suppress the unit identifier $i$ in the notation for $X$ and $\sigma^2$, 
as we are focusing on a generic unit. 
The Lagrange-Euler theorem guides us to ignore for a moment that $t$ is a function and to consider
derivatives of $F$ in $t$ as a real-valued argument:
\begin{eqnarray}
\label{eq:el1}
F_t( t, \sigma^2 ) := \frac{d}{dt} F(t, \sigma^2 ) = 0  \quad {\mbox {\rm for all $\sigma^2 $ in the support of $g$ }} .
\end{eqnarray}
This Lagrange-Euler equation simplifies, 
\begin{eqnarray*}
F_t(t,\sigma^2) &=& \frac{d}{dt} 
 	\left\{  \int_{\theta_\alpha}^\infty P(X \geq t| \theta, \sigma^2 ) \, f(\theta|\sigma^2)
 	\, d\theta  + \lambda P(X \geq t | \sigma^2 ) \right\} \\
 &=& -p(t|\sigma^2) \left\{ \int_{\theta_\alpha}^\infty \frac{ p(t|\theta, \sigma^2) f(\theta | \sigma^2) }{
		p(t|\sigma^2) } + \lambda \right\} \\
 &=& -p(t|\sigma^2) \left\{ P( \theta \geq \theta_\alpha | X=t, \sigma^2 ) + \lambda \right\}.
\end{eqnarray*}
In the above development, $p(t|\theta, \sigma^2)$ is the sampling density of $X$ given $\theta$
and $\sigma^2$ evaluated at the argument $t$, and similarly $p(t|\sigma^2)$ is the density
marginal to $\theta$ but conditional upon $\sigma^2$. 
Solving $F_t(t, \sigma^2)=0$ for all $\sigma^2>0$ gives the result~(\ref{eq:define}). 

\subsection{Theorem 2}
Let $\alpha$ and $\lambda$ both be fixed  in $(0,1)$, and for 
binary statistics $a=a(X,\sigma^2) \in \{0,1\}$ consider the objective
function:
\begin{eqnarray}
\label{eq:k1}
I_{\alpha, \lambda} (a) = 
 E\left[ a(X,\sigma^2) \left\{ 1(\theta \geq \theta_\alpha) - \lambda \right\} \right].
\end{eqnarray}
Maximizing $I_{\alpha, \lambda} (a)$ is achieved by maximizing the conditional expectation
\begin{eqnarray*}
 E\left[ \left. a(X,\sigma^2) \left\{ 1(\theta \geq \theta_\alpha) - \lambda \right\}  \right| 
 X, \sigma^2 \right]
\end{eqnarray*}
for every conditioning event, but this conditional expectation is
$a(X,\sigma^2) \left\{ V_\alpha(X,\sigma^2) - \lambda \right\}$,
which is maximized at $a^*_{\alpha, \lambda} (X,\sigma^2) = 
 1\left\{ V_\alpha(X,\sigma^2) \geq \lambda \right\}$.
Now we select a particular value $\lambda_\alpha$ of $\lambda$ for which
$E\{ a^*_{\alpha, \lambda} (X, \sigma^2) \} = \alpha$, we denote the resulting
rule by $\hat a_\alpha = a^*_{\alpha, \lambda_\alpha}$, and we construct the
threshold function:
\begin{eqnarray}
\label{eq:key}
t^*_\alpha(\sigma^2) = \inf \{ x: \hat a_\alpha(x, \sigma^2) = 1 \}
 = \inf \{ x: V_\alpha( x, \sigma^2) \geq \lambda_\alpha  \}.
\end{eqnarray}  
By right continuity and monotonicity it follows that $X \geq t^*_\alpha(\sigma^2)$ is equivalent to
$V_\alpha(X,\sigma^2) \geq \lambda_\alpha$. 
Note that the equivalence will also hold if there are values of $\sigma^{2}$ such that
$V_{\alpha}(x, \sigma^{2}) < \lambda_{\alpha}$ for all $x$ or if there are values of $\sigma^{2}$
with $V_{\alpha}(x,\sigma^{2}) \geq \lambda_{\alpha}$ for all $x$, where $t_{\alpha}^{*}( \sigma^{2} )$ is
set to positive infinity and negative infinity respectively.
 This equivalence implies the size constraint, but also
allows us to develop a comparison of the thresholds $\{ t_\alpha^* \}$ and any other
thresholds $\{ t_\alpha \}$ which also satisfy that constraint.  Using the optimality
 of $\hat a_\alpha$ in~(\ref{eq:k1}), it follows that
\begin{eqnarray}
\label{eq:k3}
I_{\alpha, \lambda_\alpha} (\hat a_\alpha)   \geq 
I_{\alpha, \lambda_\alpha} ( b_\alpha)  
\end{eqnarray}
where $b_\alpha(X,\sigma^2) = 1\left\{ X \geq t_\alpha(\sigma^2) \right\}$ is the threshold-based
rule we are comparing to the putative optimal threshold. Expanding~(\ref{eq:k3}),
\begin{eqnarray*}
 P\left\{ X \geq t^*_\alpha( \sigma^2 ), \theta \geq \theta_\alpha \right\} -
   \lambda_\alpha P\left\{ X \geq t^*_\alpha( \sigma^2 ) \right\}
\geq
 P\left\{ X \geq t_\alpha( \sigma^2 ), \theta \geq \theta_\alpha \right\} -
   \lambda_\alpha P\left\{ X \geq t_\alpha( \sigma^2 ) \right\}
\end{eqnarray*}
from which optimality of $\{ t^*_\alpha \}$ follows immediately, since both marginal probabilities 
involved equal $\alpha$.

\subsection{Theorem 3}
 Suppose there is crossing, in contradiction to the claim: i.e.  there exists 
$(\alpha_{1}, \alpha_{2}, \sigma_{0}^{2})$ with $\alpha_{1} < \alpha_{2}$ such that
$t_{\alpha_{1}}( \sigma_{0}^{2} ) = t_{\alpha_{2}}( \sigma_{0}^{2} )$.  By the mean-value theorem,
there exists $c \in [\alpha_{1},\alpha_{2}]$ such that
$\frac{ \partial t_{\alpha}(\sigma_{0}^{2}) }{\partial \alpha}\Big|_{\alpha = c}
= ( t_{\alpha_{2}}( \sigma_{0}^{2} ) - t_{\alpha_{1}}( \sigma_{0}^{2} )) /( \alpha_{2} - \alpha_{1} ) = 0$, which is in violation 
of the derivative condition.

In the normal/normal model, $t^*_\alpha(\sigma^2) = \theta_\alpha (\sigma^2 + 1) - u_\alpha \sqrt{ \sigma^2 (\sigma^2 + 1 ) }$
 as presented in~(\ref{eq:tstar}),
with $\theta_\alpha = \Phi^{-1}( 1-\alpha )$,  $u_\alpha$ defined by the constraint equation~(\ref{eq:ci}), and $\Phi$ the
 cumulative distribution function of the standard normal.
Our proof that this threshold has a negative derivative in $\alpha$
 uses the interesting fact that
$h(a) = \phi\Big[ \Phi^{-1}(a) \Big]$ is strictly concave for $a \in (0,1)$, which may be confirmed by differentiation. (Here $\phi$ 
is the density function associated with $\Phi$.)

\begin{lma}
\label{lemma:nncross}
In the normal/normal model, assuming $P\{ \sigma^{2} = 0 \} < 1$, we have 
$ \frac{d u_\alpha }{d \alpha} > \frac{d \theta_\alpha }{ d \alpha } $ . 
\end{lma}

\begin{proof}
Let 
\begin{equation}
\label{eq:five}
D_{\alpha}(\sigma^{2}) = \Phi\Big\{ \theta_{\alpha}\sqrt{\sigma^{2} + 1} - u_{\alpha}\sigma \Big\},
\end{equation}
so that $E\big[ D_{\alpha}(\sigma^{2}) \big] = 1 - \alpha$ is the constraint equation~(\ref{eq:ci}).
Suppose, by contradiction, that $-u_{\alpha}' \geq -\theta_{\alpha}'$, where primes indicate differentiation with respect to $\alpha$. 
 Differentiating~(\ref{eq:ci}) with respect to $\alpha$, and 
 using $G$ to denote the distribution function of $\sigma_i^2$,  we get:
\begin{eqnarray}
1 &=& -\theta_{\alpha}'\int_{0}^{\infty}{ \sqrt{\sigma^{2} + 1}\phi\Big\{ \Phi^{-1}\big[ D_{\alpha}(\sigma^{2}) \big] \Big\} dG(\sigma^{2})  }
- (-u_{\alpha}')\int_{0}^{\infty}{ \sigma\phi\Big\{ \Phi^{-1}\big[ D_{\alpha}(\sigma^{2}) \big] \Big\} dG(\sigma^{2})   }  \nonumber \\
&\leq& -\theta_{\alpha}'\int_{0}^{\infty}{ \sqrt{\sigma^{2} + 1}\phi\Big\{ \Phi^{-1}\big[ D_{\alpha}(\sigma^{2}) \big] \Big\} dG(\sigma^{2})  }
+ \theta_{\alpha}'\int_{0}^{\infty}{ \sigma \phi\Big\{ \Phi^{-1}\big[ D_{\alpha}(\sigma^{2}) \big] \Big\} dG(\sigma^{2})   } \nonumber \\
&=& -\theta_{\alpha}'\int_{0}^{\infty}{ \big(\sqrt{\sigma^{2} + 1} - \sigma\big)\phi\Big\{ \Phi^{-1}\big[ D_{\alpha}(\sigma^{2}) \big] \Big\} dG(\sigma^{2})  } \nonumber \\
&<& -\theta_{\alpha}'\int_{0}^{\infty}{\phi\Big\{ \Phi^{-1}\big[ D_{\alpha}(\sigma^{2}) \big] \Big\} dG(\sigma^{2})  } 
\quad \textrm{ unless } P\{\sigma^{2} = 0\} = 1 \nonumber \\
&=& -\theta_{\alpha}'E\Big\{ h\big( D_{\alpha}(\sigma^{2}) \big)  \Big\}.
\end{eqnarray}
From Jensen's inequality, we know that 
$E\Big\{ h\big( D_{\alpha}(\sigma^{2}) \big)  \Big\} \leq h\Big\{ E\big( D_{\alpha}(\sigma^{2}) \big)  \Big\}$.
Hence,
\begin{eqnarray*}
1 < -\theta_{\alpha}'h\Big\{ E\big( D_{\alpha}(\sigma^{2}) \big)  \Big\} 
= -\theta_{\alpha}'h(1 - \alpha) 
= -\theta_{\alpha}'\phi\big[ \Phi^{-1}(1 - \alpha) \big] 
= 1.
\end{eqnarray*}
This contradiction leads us to conclude that $-u_{\alpha}' < -\theta_{\alpha}'$, thus establishing the lemma.
\end{proof}

To complete the non-crossing proof, we
differentiate~(\ref{eq:tstar}) in $\alpha$: 
\begin{eqnarray*}
\frac{ \partial t^*_\alpha (\sigma^2 ) }{ \partial \alpha } 
        & = & (\sigma^2+1) \left(  \frac{ d \theta_\alpha }{ d \alpha } -  \frac{ du_\alpha }{d \alpha } \sqrt{ \frac{ \sigma^2 }{ \sigma^2 + 1 } } \right) \\
    &<& (\sigma^2 +1 ) \left( \frac{ d \theta_\alpha }{ d \alpha } -  \frac{ d\theta_\alpha }{d \alpha } \sqrt{ \frac{ \sigma^2 }{ \sigma^2 + 1 } } \right) \\
    &=& \frac{d \theta_\alpha }{ d \alpha } (\sigma^2 + 1) \left( 1 - \sqrt{ \frac{ \sigma^2 }{ \sigma^2 + 1 } } \right) 
    < 0,
\end{eqnarray*}
where the first inequality comes from {\sc Lemma}~1 and the second from the fact that
  $\frac{d \theta_\alpha }{ d \alpha } = -1/\phi(\theta_\alpha) < 0 $ .
For the trivial case when $P\{ \sigma^{2} = 0 \} = 1$, we 
 note that the optimal ``threshold function"
is $t_{\alpha}^{*}(0) = \theta_{\alpha}$ which obviously satisfies $\frac{\partial t_{\alpha}^{*}(\sigma^{2})}{\partial \alpha} < 0$.

\subsection{Theorem 4}
We proceed in steps.
\begin{lma}
\label{lemma:tconverge}
Assume A1 and A2. For each $\alpha \in (0,1)$, $\hat \theta_{\alpha,n}
 \longrightarrow_{a.s.} \theta_\alpha$ as $n \longrightarrow \infty$.
\end{lma}
\begin{proof}
 At continuity points $p$ of $F^{-1}$, 
  $\hat F_n^{-1}(p)$ converges almost
 surely to the limiting quantile
 $F^{-1}(p)$ by A2 and, for example, Lemma~21.2 of van der Vaart (1998).
 Continuity of $F^{-1}$ follows from~A1, and thus the result follows.
\end{proof}

\begin{lma}
\label{lemma:vconverge}
Assume A1-A3. The limiting posterior tail probability $V_\alpha(X_i, \sigma_i^2)$ is continuous
 and nondecreasing in $\alpha$ for any data $(X_i,\sigma_i^2)$. Further, as $n \longrightarrow \infty$, 
\begin{eqnarray*}
  \sup_{ \alpha \in (0,1) } \left| \hat V_{\alpha,n}(X_i, \sigma_i^2 ) - V_{\alpha}(X_i, \sigma_i^2 ) \right|
  \longrightarrow_{a.s.} 0.
\end{eqnarray*}
\end{lma}

\begin{proof}
First we confirm pointwise (in $\alpha$) convergence of the numerator and the denominator
 of~(\ref{eq:vhat}) when evaluated at $(X_i, \sigma_i^2)=(x,\sigma^2)$. 
 The denominator
 is immediate, owing to $p(x|\theta,\sigma^2)$ being bounded and continuous
  in $\theta$, and owing to the almost sure weak convergence of $\hat F_n$.  
For the numerator, note that the mapping $\theta \mapsto 
 1( \theta \geq \theta_\alpha ) \, p(x|\theta,\sigma^2 )$, for fixed 
 $(x, \sigma^2)$, is continuous except at 
 $\theta_\alpha$, which  has zero point 
 mass in the limiting distribution $F$. Thus 
  $\int_{\theta_\alpha}^\infty p(x| \theta, \sigma^2 ) 
 \, d\hat F_n(\theta) $ converges almost surely 
 to $\int_{\theta_\alpha}^\infty p(x|\theta, \sigma^2) \, dF(\theta) $,
  using A2 and, for example, Theorem~2.3 of van der Vaart (1998). 
 It is sufficient to confirm that the error $e_n$, defined
 $e_n = \Big| \int_{\hat \theta_{\alpha,n}}^\infty 
 p(x|\theta,\sigma^2) \, 
  d\hat F_n(\theta)  - \int_{\theta_\alpha}^\infty p(x|\theta,\sigma^2) \, 
  d \hat F_n(\theta)
 \Big|$ converges almost surely to zero. With the bound~A3(ii),
  and taking any $\epsilon >0 $, 
 we have 
\begin{eqnarray*}
 e_n &\leq&  K(\sigma^2)
    \int_{\mathbb R} 
 \Big| 1( \theta \geq \hat \theta_{\alpha, n} ) - 1( \theta \geq \theta_\alpha)
 \Big| \, d\hat F_n(\theta)  
 \leq  K(\sigma^2)  \int_{\theta_\alpha - \epsilon}^{\theta_\alpha + \epsilon}
 \, d\hat F_n(\theta)  \qquad {\mbox {\rm for $n \geq N_\epsilon$ } },  
\end{eqnarray*}
where the second inequality is almost sure owing to {\sc Lemma}~\ref{lemma:tconverge}.
Consequently, $\limsup_n e_n $ is almost surely bounded by
  $ K(\sigma^2) \left\{ F(\theta_\alpha+\epsilon)
   - F(\theta_\alpha - \epsilon ) \right\}$ for every $\epsilon>0$, and
 so, pointwise in $\alpha$, the limiting error must be zero, since $F$ contains no atoms.
On continuity and monotonicity of $\alpha \mapsto V_\alpha(D_i)$, let $\{\alpha_n\}$
 denote a sequence in $(0,1)$ for which $\alpha_n \geq \alpha$.
We have
\begin{eqnarray*}
0 \leq V_{\alpha_n}(D_i) - V_\alpha(D_i) 
   &=& \frac{1}{p(D_i)} \int_{\theta_{\alpha_n}}^{\theta_\alpha} p(X_i | \theta_i, 
 		\sigma_i^2 ) \, dF(\theta_i) \\
   &\leq& \frac{1}{p(D_i)} K(\sigma_i^2) (\alpha_n - \alpha) .
\end{eqnarray*}
 Monotonicity is immediate from this, but
also, if $\alpha_n \longrightarrow \alpha$,
 we obtain right continuity of $V_{\alpha_n}(D_i)$.  A comparable argument gives
 left continuity.
Uniform convergence follows from Polya's theorem (e.g., Bickel and Millar, 1992).
\end{proof}

\begin{lma}
\label{lemma:vprops}
If A3, the mappings $(x, \sigma^2) \mapsto V_\alpha(x, \sigma^2)$ and $(x, \sigma^2) \mapsto \hat V_{\alpha,n}(x, \sigma^2)$
 are continuous. Further, for any $x_1>x_0$, $V_\alpha(x_1, \sigma^2) > V_\alpha(x_0, \sigma^2)$ 
  and $\hat V_{\alpha, n}(x_1, \sigma^2) > \hat V_{\alpha,n}( x_0, \sigma^2)$.
\end{lma}
 \begin{proof}
Take a sequence $\{d_m=(x_m,\sigma^2_m)\}$ with 
 $d_m \longrightarrow d=(x,\sigma^2)$, and observe that for each $n$,
 \begin{eqnarray*}
 \lim_{m \longrightarrow \infty} \int_{-\infty}^{\infty} p(x_m|\theta, 
 \sigma_m^2) \, d\hat F_n(\theta)
 = \int_{-\infty}^\infty \lim_{m \longrightarrow \infty}  p(x_m|\theta, \sigma^2_m) \,
 d \hat F_n( \theta ) 
 =  \int_{-\infty}^\infty p(x|\theta, \sigma^2) \, d \hat F_n( \theta ). 
 \end{eqnarray*} 
 The first equality follows from a dominated convergence argument, using A3(ii), and
 the second equality follows from continuity of the local sampling density, A3(i).
 The same would hold if we replaced the integrand 
 $p(x_m|\theta, \sigma^2_m)$ with 
 $1( \theta \geq \hat \theta_{\alpha,n} ) p(x_m | \theta, \sigma^2_m )$,
 and likewise modified the limit.
 Thus continuity of the ratio $\hat V_{\alpha,n}(x,\sigma^2)$ is established. 
 The argument for $V_\alpha(x,\sigma^2)$ is analogous.

 On the monotonicity claim, note that
$V_{\alpha}(x,\sigma^{2}) = 1/\{1 + 1/\psi(x) \}$,
where 
\begin{eqnarray*}
\psi(x) = \int_{\theta_{\alpha}}^{\infty}{ p(x|\theta,\sigma^{2}) \, dF(\theta)}  \Big/
\int_{-\infty}^{\theta_{\alpha}}{ p(x|\theta,\sigma^{2})  \, dF(\theta)} .
\end{eqnarray*}
Showing that $\psi(x)$ is increasing would be enough to prove $V_\alpha(x,\sigma^2)$ is increasing.
 Write
\begin{equation}
\frac{\psi(x_{1})}{\psi(x_{0})} =
\left( \frac{ \int_{\theta_{\alpha}}^{\infty}{ p(x_{1}|\theta,\sigma^{2}) \, dF(\theta)} }
{ \int_{\theta_{\alpha}}^{\infty}{ p(x_{0}|\theta,\sigma^{2}) \, dF(\theta)} } \right)
\left(\frac{\int_{-\infty}^{\theta_{\alpha}}{ p(x_{0}|\theta,\sigma^{2}) \, dF(\theta)} }
{ \int_{-\infty}^{\theta_{\alpha}}{ p(x_{1}|\theta,\sigma^{2})  \, dF(\theta)} } \right)
= y_{1}y_{2}.
\label{eq:bayes_factor}
\end{equation}
If we let $\rho(\theta) = p(x_{1}|\theta,\sigma^{2})/p(x_{0}|\theta,\sigma^{2})$, then,
because $\rho(\theta)$ is increasing by A3:
\begin{eqnarray*}
y_{1} = \frac{ \int_{\theta_{\alpha}}^{\infty}{ p(x_{1}|\theta,\sigma^{2}) \, dF(\theta)} }
{ \int_{\theta_{\alpha}}^{\infty}{ p(x_{0}|\theta,\sigma^{2}) \, dF(\theta)} }
 &=&  \frac{ \int_{\theta_{\alpha}}^{\infty}{ p(x_{0}|\theta,\sigma^{2}) \rho(\theta)  \,  dF(\theta)} }
{ \int_{\theta_{\alpha}}^{\infty}{ p(x_{0}|\theta,\sigma^{2}) \, dF(\theta)} } \\
&>& \frac{ \rho(\theta_{\alpha})\int_{\theta_{\alpha}}^{\infty}{ p(x_{0}|\theta,\sigma^{2}) \,  dF(\theta)} }
{ \int_{\theta_{\alpha}}^{\infty}{ p(x_{0}|\theta,\sigma^{2}) \, dF(\theta)} }
= \rho(\theta_{\alpha}).
\end{eqnarray*}
Likewise,
\begin{eqnarray*}
y_{2} = \frac{\int_{-\infty}^{\theta_{\alpha}}{ p(x_{0}|\theta,\sigma^{2}) \, dF(\theta)} }
{ \int_{-\infty}^{\theta_{\alpha}}{ p(x_{1}|\theta,\sigma^{2}) \,  dF(\theta)} } 
&=& \frac{\int_{-\infty}^{\theta_{\alpha}}{ p(x_{0}|\theta,\sigma^{2})  \, dF(\theta)} }
{ \int_{-\infty}^{\theta_{\alpha}}{ p(x_{0}|\theta,\sigma^{2})\rho(\theta) \, dF(\theta)} }  \\
& > &\frac{\int_{-\infty}^{\theta_{\alpha}}{ p(x_{0}|\theta,\sigma^{2})  \, dF(\theta)} }
{ \rho(\theta_{\alpha})\int_{-\infty}^{\theta_{\alpha}}{ p(x_{0}|\theta,\sigma^{2}) \, dF(\theta)} } 
= \frac{1}{\rho(\theta_{\alpha})}.
\end{eqnarray*}
Hence, $y_{1}y_{2} > 1$ and from (\ref{eq:bayes_factor})
we know that $\psi(x_{1})/\psi(x_{0}) > 1$.  Whence $\psi(x)$ is increasing.
The argument for monotonicity of $\hat V_{\alpha,n}(x, \sigma^2) $ is completely analogous.
\end{proof}

\begin{lma}
\label{lemma:hconverge}
Let $B = (0,1) \times (0,1)$ denote the open unit square.
Assume A1-A3, and let $\hat H^{-1}_{\alpha, n}(p)$ be the quantile function associated with the empirical distribution of
 $\{ \hat V_{\alpha,n}(X_i, \sigma_i^2) \}$, for some $(\alpha, p) \in B$.
 As $n \longrightarrow \infty$, $\hat H_{\alpha,n}^{-1}(p) \longrightarrow_P H_\alpha^{-1}(p)$, and this limit is
 continuous on $B$. The limit function and each estimate are 
 nondecreasing in $p$ for each $\alpha$ and nondecreasing in $\alpha$
 for each $p$. Furthermore, the convergence is uniform on any closed square
 $A_\delta = [\delta, 1-\delta] \times [\delta, 1-\delta]$ for $\delta \in (0,1/2) $.
\end{lma}
\begin{proof}
To simplify notation,
let $\xi_i = V_{\alpha}(X_{i},\sigma_{i}^{2})$ and $\hat \xi_{n,i}
= \hat{V}_{\alpha,n}(X_{i},\sigma_{i}^{2})$.
For $v\in (0,1)$, we consider the intermediate empirical distribution
$\tilde H_{\alpha,n}(v) = \frac{1}{n} \sum_{i=1}^n 1\left[ \xi_i \leq v
 \right]$,
which entails no estimation error in $\hat F_n$ as compared to the
 computable $\hat H_{\alpha,n}(v) = \frac{1}{n} 
 \sum_{i=1}^n 1( \hat \xi_{n,i} \leq v )$, and which
converges to $H_\alpha(v)$ by the law of large numbers, owing to A1. 
To show that $\hat H_{\alpha,n}$ converges, further define
$ \Delta_n = \Big| \hat{H}_{\alpha,n}(v) - \tilde{H}_{\alpha,n}(v) \Big| $.
For any $\epsilon > 0$ we have
 \begin{eqnarray*}
 \Delta_n   
  & \leq & \frac{1}{n} \sum_{i=1}^n \Big| 
 1[ \hat \xi_{n,i} \leq v ] -  1[ \xi_i \leq v ] \Big|  \\
 &=& \frac{1}{n} \sum_{i=1}^n \Big| 
 1[ \hat \xi_{n,i} \leq v ] -  1[ \xi_i \leq v ] \Big| (1-U_{n,i})  +
 \frac{1}{n}\sum_{i=1}^n \Big| 1[ \hat \xi_{n,i} \leq v ] -  1[ \xi_i \leq v ] \Big| U_{n,i} 
\end{eqnarray*}
where $U_{n,i} = 1( |\xi_i - \hat \xi_{n,i} | > \epsilon )$.
Thus
\begin{eqnarray*}
\Delta_n &\leq& \frac{1}{n} \sum_{i=1}^n \Big|      
 1[ \hat \xi_{n,i} \leq v ] -  1[ \xi_i \leq v ] \Big| (1-U_{n,i})  +
  \frac{1}{n} \sum_{i=1}^n U_{n,i} \\
 &\leq & \frac{1}{n} \sum_{i=1}^n 1\{ \hat \xi_{n,i} \in (v, v+\epsilon] \}
        + \frac{1}{n} \sum_{i=1}^n 1\{ \xi_i \in (v, v+\epsilon ] \} 
 + \frac{1}{n} \sum_{i=1}^n U_{n,i}.
\end{eqnarray*}
From the symmetry of $\hat F_n$ in A2, 
 all $\hat \xi_{n,i}$ are identically distributed, and hence taking
 expectations,
\begin{eqnarray}
\label{eq:edelta}
E(\Delta_n) \leq P\{ \xi_{n,1} \in (v, v+\epsilon] \} +
                 P\{ \xi_1 \in (v, v+\epsilon] \} + E(U_{n,1}).
\end{eqnarray}
As $n \longrightarrow \infty$, the term $E(U_{n,1})$ converges to $0$ by 
 {\sc Lemma}~\ref{lemma:vconverge}, and likewise the upper bound in~(\ref{eq:edelta})
 converges to 
 $2 P\{ \xi_1 \in (v, v+\epsilon] \}=2\left[ H_\alpha(v+\epsilon) - H_\alpha(v) \right]$. 
 Because $\epsilon>0$ could be arbitrarily small, and using the 
 continuity
 of $H_\alpha$ (see {\sc Lemma}~S1, 
 Supplementary Material), it follows that $\Delta_n \longrightarrow_P 0$ and hence
 $\hat H_{\alpha,n}(v) \longrightarrow_P H_\alpha(v)$ as $n \longrightarrow \infty$.
 Convergence in probability of $\hat H^{-1}_{\alpha, n}(p)$
 to $H^{-1}_\alpha(p)$ follows from a basic fact about
 distributions (see {\sc Lemma} S2, Supplementary Material).  Continuity of the limit $H_\alpha^{-1}(p)$ on $B$ 
 and coordinatewise monotonicity
 follow from the model regularity conditions (see {\sc Lemma} S1, Supplementary Material).  Interestingly,
 there are two discontinuities on the closed square, at $(0,1)$ and $(1,0)$, where the function switches 
 immediately between zero and one. 
 Thus we avoid having $\alpha$ near the boundary in establishing uniformity 
 of convergence, which itself follows from a 2-d version of Polya's theorem, owing to 
 coordinatewise monotonicity and continuity of the limit (see {\sc Lemma} S3, Supplementary
 Material).
\end{proof}

\begin{lma}
\label{lemma:lconverge}
Define $g_\alpha(D_i) = V_\alpha(D_i) - \lambda_\alpha$ and $\hat g_{\alpha,n}(D_i) = \hat V_{\alpha,n}(D_i) - \hat \lambda_{\alpha,n}$,
  and assume A1-A4.  Both
 $\sup_{\alpha \in [\delta,1-\delta]} | \hat \lambda_{\alpha,n} - \lambda_\alpha |$ 
 and $\sup_{\alpha \in [\delta,1-\delta]} \left| \hat g_{\alpha,n}(D_i) - g_\alpha(D_i)  
   \right| $ converge to zero in probability as $n \longrightarrow \infty$, for any fixed $\delta \in (0,1/2)$.
 Further, $\alpha \mapsto g_\alpha(D_i)$ is continuous and $g_\alpha(D_i)=0$
 has a unique root $r(D_i)$.
\end{lma}
\begin{proof}
Let $A_\delta = [\delta, 1-\delta] \times [\delta, 1-\delta]$ denote a closed square
 within $B$, and note that:
\begin{eqnarray*}
\sup_{\alpha \in [\delta, 1-\delta]} \left| \hat \lambda_{\alpha,n} - \lambda_\alpha
  \right| = \sup_{\alpha \in [\delta,1-\delta]} \left| \hat H_{\alpha,n}^{-1}(1-\alpha)
 - H_\alpha^{-1}(1-\alpha) \right| \leq \sup_{ (\alpha, p) \in A_\delta }
  \left| \hat H_{\alpha,n}^{-1}(p) - H_{\alpha}^{-1}(p) \right|.
\end{eqnarray*}
Uniform convergence of $\hat \lambda_{\alpha,n}$ follows from 
 {\sc Lemma}~\ref{lemma:hconverge}.  Similarly, uniform convergence of 
 $\hat g_{\alpha,n}(D_i)$ follows  after also invoking {\sc Lemma}~\ref{lemma:vconverge}.

 Continuity of $g_\alpha(D_i)$ follows from {\sc Lemmas}~\ref{lemma:vconverge} 
 and~\ref{lemma:hconverge}.
 We deduce uniqueness in the $\alpha$ root of $g_\alpha(D_i)=0$ first by noting that A1 and
 continuity of $V_\alpha$
 in data ({\sc Lemma}~\ref{lemma:vprops}) 
 imply the existence of $\lambda_\alpha$ satisfying~(\ref{eq:lambda}).
 Were there not at least one $\alpha$ value for which $g_\alpha(D_i)=0$,
 then either $V_\alpha(X_i,\sigma_i^2)$ would always exceed $\lambda_\alpha$ or would always be dominated by it.  Take the second case; the first is analogous.
 Find an open ball around $D_i = (X_i, \sigma_i^2)$  
 such that $g_\alpha(d) < 0$
    for all $d$ in this ball and for all $\alpha$.  This ball has some positive
 probability, say $\epsilon>0$, and so $P[ V_\alpha(D_i) < \lambda_\alpha ] \geq \epsilon$.
 But owing to~(\ref{eq:lambda}), we have a contradiction when $\alpha > 1-\epsilon$, implying
 that there must be at least one root of $g_\alpha(D_i) = 0$.  The conditions of Theorem~2
are met, and so from continuity in {\sc Lemma}~\ref{lemma:vprops}, $V_\alpha[ t^*_\alpha(\sigma_i^2), \sigma_i^2]
 = \lambda_\alpha$ defines the optimal threshold.  Finally, by A4, 
 $X_i = t^*_\alpha(\sigma_i^2)$ at exactly one  value of $\alpha$.
\end{proof}

\begin{lma}
\label{lemma:rconverge}
 If A1-A4, then for $\delta \in (0,1/2)$, $\hat r_n^\delta(D_i) \longrightarrow_P r^\delta(D_i)$ as $n \longrightarrow \infty$.
\end{lma}

\begin{proof}
The empirical r-value $\hat r_n^\delta(D_i)$ is like a root of $\hat g_{\alpha,n}(D_i) = 0$, at least truncated
away from endpoints $0$ and $1$, but owing to the sample quantile estimation, $\hat g_{\alpha,n}(D_i)$ is not continuous at all $\alpha$
and may admit multiple roots.  In spite of this, {\sc Lemma}~\ref{lemma:lconverge} assures not only continuity of the limit $g_\alpha(D_i)$ having
 a unique root $r(D_i)$, but also uniform convergence of sample functions $\hat g_{\alpha,n}(D_i)$ to this limit, at least on compact 
subsets of $(0,1)$.   From the first of these properties, the extreme value theorem implies that 
for any sufficiently small $\epsilon > 0$, there exists $\nu=\nu(D_i)>0$ such that 
 the limit function $g_\alpha(D_i)$ has magnitude at least $\nu$ for all $\alpha$ with $|r(D_i) - \alpha| > \epsilon$.
From the uniform convergence,  $\left| \hat g_{\alpha, n}(D_i) - g_\alpha(D_i) \right| < \nu/2 $ with high probability 
 for large $n$, uniformly for $\alpha \in [\delta, 1-\delta]$, and thus in this event 
 $|\hat r^\delta_n(D_i) - r(D_i)|  < \epsilon $.
 {\sc Lemma}~S4, Supplementary Material, provides further details. 
\end{proof}
Proceeding to prove Theorem~4, we know from the unique-root result in {\sc Lemma}~\ref{lemma:lconverge}
that events $[r(D_i) \leq \alpha]$ and $[V_\alpha(D_i) \geq \lambda_\alpha]$ are equivalent,
and so the ideal $r(D_i)$ has a Uniform$(0,1)$ distribution by~(\ref{eq:lambda}). The first claim follows from 
 {\sc Lemma}~\ref{lemma:rconverge} and, for example, Theorem 2.3 from van der Vaart (1998), using the fact
 that $1[ r^\delta(D_i) \leq \alpha ] = 1[ r(D_i) \leq \alpha]$ for $\alpha \in [\delta,1-\delta]$. 
  R-values in~(\ref{eq:converge2}) 
 do not involve truncation away from endpoints $0$ or $1$.  The claimed 
 lower bound $A_\alpha := P[ r(D_i) \leq \alpha , \theta_i \geq \theta_\alpha]$
 is maximal because 
 the conditions of Theorem~2 are satisfied ({\sc Lemma}~\ref{lemma:vprops}),
 and because the maximal agreement is achieved using $r(D_i)$ (unique-root remarks above).  
 To establish the bound, let $\hat A_{\alpha,n}$
 denote the left hand side of~(\ref{eq:converge2}), and introduce $\tilde A_{\alpha,n} = \frac{1}{n}
 \sum_{i=1}^n 1[ r(D_i) \leq \alpha , \theta_i \geq \theta_\alpha]$.  Of course $\tilde A_{\alpha,n} \longrightarrow_P A_\alpha$
 by the law of large numbers, so at issue are deviations between $\hat A_{\alpha,n}$ and $\tilde A_{\alpha,n}$ caused by estimation
 errors.  With $\alpha \in [\delta,1-\delta]$, $\hat r_n^\delta(D_i) \leq \alpha$ 
 implies that $\hat r_n(D_i) \leq \alpha$,  and therefore
 $\hat A_{\alpha,n} \geq \frac{1}{n} \sum_{i=1}^n 1\left[ \hat r_n^\delta(D_i) \leq \alpha, \theta_i \geq \theta_\alpha \right]$. 
 Now decompose this lower bound into $\tilde A_{\alpha,n} + e_{n}$, where, 
$ e_{n} = \frac{1}{n} \sum_{i=1}^n \left\{  1[ \hat r^\delta_n(D_i) \leq \alpha ] 
 					- 1[ r(D_i) \leq \alpha ] \right\} 1[ \theta_i \geq \theta_\alpha ]$,
 using the fact that $1[r^\delta(D_i) \leq \alpha]  = 1[r(D_i) \leq \alpha]$ for $\alpha \in [\delta, 1-\delta]$. 
Having convergence of $e_{n}$  in probability to zero would complete the proof.
 We have 
 \begin{eqnarray*}
 | e_{n} | \leq \frac{1}{n} \sum_{i=1}^n 1( \theta_i \geq \theta_\alpha ) \, 
  \Big| 1[  \hat r^\delta_n(D_i) \leq \alpha ]
        - 1[ r^\delta(D_i) \leq \alpha ]  \Big|.
\end{eqnarray*}
By the identical distribution of terms, induced by permutation invariance (A2),
\begin{eqnarray*}
 E|e_{n}| &\leq& E\left\{ 1(\theta_i \geq \theta_\alpha ) \,
  \Big|  1[  \hat r^\delta_n(D_1) \leq \alpha ]
        - 1[ r^\delta(D_1) \leq \alpha ]  \Big| \right\} \\
 &\leq & \sqrt{\alpha} 
 \sqrt{
 E \Big| 1[  \hat r^\delta_n(D_1) \leq \alpha ]
        - 1[ r^\delta(D_1) \leq \alpha ]   \Big| },
\end{eqnarray*}
with the second inequality by Cauchy-Schwartz.
  The integrand within the expectation on the right hand side is bounded by 1 and converges in probability to 0 
 ({\sc Lemma}~\ref{lemma:rconverge} and Theorem 2.3, van der Vaart, 1998),
and so $e_{n} \longrightarrow_P 0$, completing the proof.

\section*{Acknowledgements}
This research was supported in part by 
grants from the US National Institutes of Health: R21 HG006568,
T32 GM074904, and U54AI117924. 
 The authors thank Christina Kendziorski and several anonymous reviewers for critical comments 
 on earlier drafts. 
Additional details on data analyses, threshold functions, and computation 
 are provided in a Supplementary Material document.

\nocite{gelman1999all,berger1988bayesian,RSSC:RSSC760,gibbons1979introduction,laird1989empirical,lin2006loss,shen1998triple,lehmann1986testing,noma2010bayesian,wright2003loss,morris2012large,hao2013limited,de2013whole,normand1997statistical,RSSA:RSSA486,hall2010modeling,xie09,mccarthy2009testing,efron2010large,niem2010evaluating,storey2003,pyeon2007fundamental,bf95,jost,coelho2007distribution,smyth2004linear,kendziorski2003parametric,leng2013ebseq,zorn,espn,lindsay,VdVaart:1998,bickel}
\bibliographystyle{Chicago}
\bibliography{rv}


\begin{table}
\caption{Threshold functions associated with various ranking criteria, normal/normal
 model  \label{tab:1} }
\centering
\fbox{%
\begin{tabular}{lcc}
criteria & ranking variable & threshold function $t_\alpha(\sigma^2) $ \\ \hline
MLE & $X_i$  & $u_\alpha$ \\
PV  $H_0: \theta_i=0$ &  $X_i/\sigma_i$  &  $u_\alpha \sigma $ \\
PV  $H_0: \theta_i=c$ &  $(X_i-c)/\sigma_i$  &  $c+ u_\alpha \sigma  $ \\
PM  &$X_i/(\sigma_i^2+1)$  & $u_\alpha (\sigma^2 + 1 ) $ \\
PER  & $P(\theta_i \leq \theta | X_i, \sigma_i^2 )$   & $u_\alpha \sqrt{ (\sigma^2+1) (2\sigma^2 +1 ) } $ \\
BF & $1(X_i>0) \frac{ P(X_i|\sigma_i^2, \theta_i \neq 0 ) }{
                P(X_i|\sigma_i^2, \theta_i=0 ) }  $ & $
        \sqrt{ \sigma^2 (\sigma^2 + 1 )
        \left\{ u_\alpha + \log \frac{ (\sigma^2 +1) }
                { \sigma^2 }  \right\} } $ \\
max agreement & r-value &  $\theta_\alpha (\sigma^2 +1 ) - u_\alpha \sqrt{ \sigma^2 (\sigma^2 + 1 ) } $  \\
\end{tabular} }
\end{table}

\begin{table}
\caption{Leading free-throw shooters, 2013-2014 regular season of the National Basketball Association.
 From $n=461$ players who attempted at least one free throw, shown are the top 25 players as inferred by r-value.
 Data $D_i$ on player $i$ include 
the number of made free throws $y_i$ and the number of attempts $m_i$. Other columns indicate
  free-throw percentage FTP$=y_i/m_i$,
 which is the maximum likelihood estimate (MLE) of the underlying ability $\theta_i$;
 posterior mean $E(\theta_i|D_i)$, r-value $\inf\{ \alpha \geq 1/n: P( \theta_i \geq \theta_\alpha | D_i) \geq \lambda_\alpha \}$; 
 qualified rank, Q.R, which is the rank of FTP amongst players for whom $y_i \geq 125$; and ranks associated with the MLE,
 posterior mean, and r-value.
 \label{tab:NBA} }
\centering
\fbox{%
\small
\begin{tabular}{lrrrrrrrrr}
 player $i$ & $y_i$ & $m_i$ & FTP & PM & RV & Q.R & MLE.R & PM.R & RV.R \\ \hline
Brian Roberts & 125 & 133 & 0.940 & 0.913 & 0.002 & 1 & 17 & 1 & 1 \\
Ryan Anderson & 59  & 62  & 0.952 & 0.898 & 0.003 &   & 15 & 2 & 2 \\
Danny Granger & 63  & 67  & 0.940 & 0.893 & 0.005 &   & 16 & 3 & 3  \\
Kyle Korver   & 87  & 94  & 0.926 & 0.892 & 0.008 &   & 19 & 4 & 4  \\
Mike Harris   & 26  & 27  & 0.963 & 0.866 & 0.010 &   & 14 & 15 & 5 \\
J.J. Redick   & 97  & 106 & 0.915 & 0.886 & 0.011 &   & 22 & 6 & 6 \\
Ray Allen     & 105 & 116 & 0.905 & 0.880 & 0.016 &   & 25 & 8 & 7 \\
Mike Muscala  & 14  & 14  & 1.000 & 0.844 & 0.017 &   & 7 & 34 & 8 \\
Dirk Nowitzki & 338 & 376 & 0.899 & 0.891 & 0.018 & 2 & 30 & 5 & 9 \\
Trey Burke    & 102 & 113 & 0.903 & 0.877 & 0.018 &   & 28 & 9 & 10 \\
Reggie Jackson& 158 & 177 & 0.893 & 0.877 & 0.024 & 3 & 32 & 11 & 11 \\
Kevin Martin  & 303 & 340 & 0.891 & 0.882 & 0.025 & 4 & 33 & 7 & 12 \\
Gary Neal     & 94  & 105 & 0.895 & 0.869 & 0.025 &   & 31 & 14 & 13 \\
D.J. Augustin & 201 & 227 & 0.885 & 0.873 & 0.031 & 5 & 38 & 12 & 14 \\
Stephen Curry & 308 & 348 & 0.885 & 0.877 & 0.031 & 6 & 39 & 10 & 15 \\
Patty Mills   & 73  & 82  & 0.890 & 0.860 & 0.032 &   & 34 & 19 & 16 \\
Courtney Lee  & 99  & 112 & 0.884 & 0.861 & 0.035 &   & 40 & 18 & 17 \\
Steve Nash    & 22  & 24  & 0.917 & 0.834 & 0.039 &   & 20.5 & 44 & 18 \\
Greivis Vasquez & 95& 108 & 0.880 & 0.857 & 0.040 &   & 41 & 22 & 19 \\
Robbie Hummel & 15  & 16  & 0.938 & 0.825 & 0.043 &   & 18 & 55 & 20 \\
Mo Williams   & 78  & 89  & 0.876 & 0.850 & 0.046 &   & 42 & 24 & 21 \\
Kevin Durant  & 703 & 805 & 0.873 & 0.870 & 0.048 & 7 & 45 & 13 & 22 \\
Aaron Brooks  & 83  & 95  & 0.874 & 0.850 & 0.049 &   & 44 & 26 & 23 \\
Damian Lillard& 371 & 426 & 0.871 & 0.865 & 0.050 & 8 & 47 & 16 & 24 \\
Nando de Colo & 31  & 35  & 0.886 & 0.831 & 0.057 &   & 37 & 48 & 25 \\ \hline
\end{tabular} }
\end{table}

\end{document}